\documentclass[final,3p,times,twocolumn]{elsarticle}

\usepackage{lineno,hyperref}

\journal{Nuclear Instruments and Methods in Physics Research Section A}

\usepackage{amsmath}  
\usepackage{amsfonts} 
\usepackage{graphicx} 
\usepackage{siunitx}
\usepackage{wasysym}
\usepackage[utf8]{inputenc}
\usepackage{lipsum}
\usepackage{booktabs}
\usepackage{stfloats}
\usepackage{caption}
\usepackage{stfloats}

\begin{document}

\begin{frontmatter}

\title{Hole Misalignment and Gain Performance of Gaseous Electron Multipliers}

\author[HIP]{Erik Br\"ucken\corref{mycorrespondingauthor}}
\ead{erik.brucken@iki.fi}
\author[HIP]{Jouni Heino}
\author[STUK]{Timo Hild\'en \corref{mycorrespondingauthor}}
\ead{timo.hilden@helsinki.fi}
\author[HIP]{Matti Kalliokoski}
\author[HIP]{Vladyslav Litichevskyi}
\author[HIP]{Raimo Turpeinen}
\author[RCP]{Dezs\H{o} Varga}

\cortext[mycorrespondingauthor]{Corresponding author}

\address[HIP]{Helsinki Institute of Physics, P.O. Box 64, FI-00014 University of Helsinki, Finland}
\address[STUK]{Radiation and Nuclear Safety Authority, P.O. Box  14, FI-00811 Helsinki, Finland}
\address[RCP]{Wigner Research Centre for Physics, 29-33 Konkoly-Thege Miklo's Str. H-1121 Budapest, Hungary}

\begin{abstract}
It is well known and has been shown that the gain performance of Gaseous Electron Multipliers (GEM) depends on the size of the holes. With an optical scanner it is possible to measure the dimensions of the holes, and to predict the performance of GEMs. However, the gain prediction of GEMs that are manufactured with a double mask etching technique is not straightforward. With the hole size information alone, it is not possible to make precise prediction of the gain. We show that the alignment of the photo-masks between the two sides of the GEM foils plays a crucial role. A misalignment of a few microns can lower the gain substantially. The study is performed by using the Helsinki high definition optical scanner for quality control of GEM foils, and this will show its true potential.
\end{abstract}


\end{frontmatter}

\section{Introduction}

\begin{figure}[h]
\centering
\includegraphics[width=0.34\textwidth,clip]{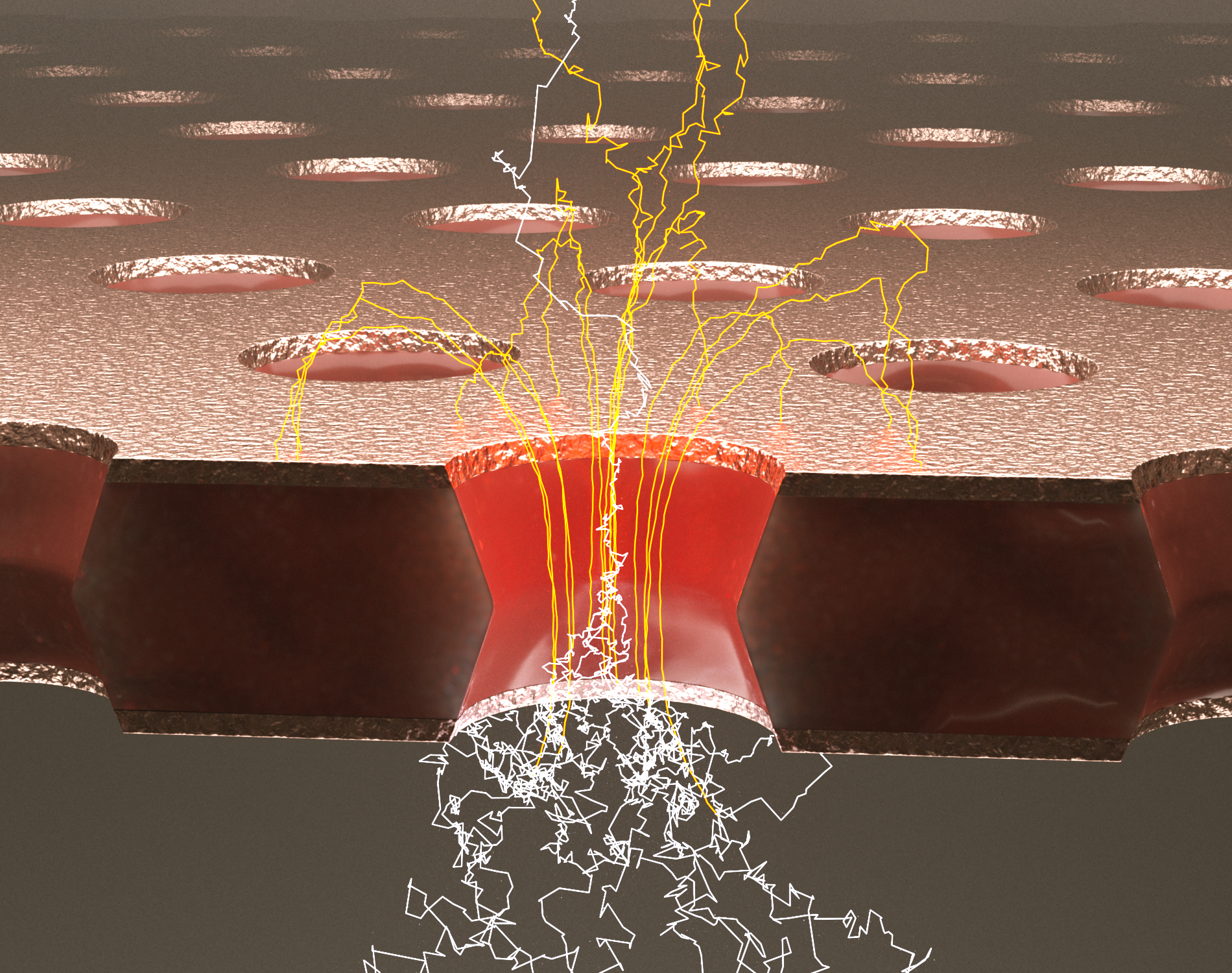}
\caption{\texttt{Garfield++} simulation of an electron avalanche in a GEM foil \cite{Veenhof,garfieldpp}. The visualization was done with \texttt{Blender}\,\cite{blender}.}
\label{fig-1}
\end{figure}

Gaseous Electron Multipliers (GEM) \cite{sauli} are gas-filled radiation detectors that are currently used in several high energy particle physics experiments. The key element is a thin polyimide foil of a 50\,$\mu$m thickness, coated with a 5\,$\mu$m thick copper layer on both sides. An array of bi-conical holes are chemically etched in a precise patterned structure with typical distances (pitch) of 140\,$\mu$m (see illustration in Fig.\,\ref{fig-1}).
The foils are placed in a gas tight enclosure filled with a gas mixture, consisting of an inert gas and a gas of polyatomic molecules. A voltage of around 300\,V to 500\,V is applied over the two copper electrodes to allow fields large enough for gas amplification. When the electrons, released by ionizing radiation in the gas, drift into the holes, Townsend avalanches, multiplying the number of drift electrons, occur. The gain of the detector is the mean number of electrons at the bottom of the foil after the avalanche in comparison to the released electrons. 

The holes in the GEM foils have typically an outer diameter of 70\,$\mu$m (copper hole) and inner diameter of~50\,$\mu$m (polyimide hole). Two types of manufacturing techniques are used based on photolithography~\cite{Bachmann}; a~double mask technique where photo-masks are on each side separately applied, and at present, more commonly used, the single mask technique where only one side is etched after applying the photo-mask. Here, the other side is electro-etched. The advantage of the latter technique is that no misalignment of the photo masks occur, but the downside is that the bi-conical shape of the holes might not be symmetric~\cite{merlin}. In this paper we study three double-mask foils with dimensions  \mbox{of 10 $\times$ 10 cm$^2$}. 

GEM based detectors show good cost efficiency for covering large areas/volumes. A recent example is the upgrade of the Time Projection Chamber of the ALICE experiment at CERN~ \cite{TPCTDR}. In such experiment it is important to know the quality of the GEM foils before the detector is assembled. Therefore a thorough quality assurance (QA) effort was undertaken \cite{erikphilly}. One important QA method was the optical scanning and gain prediction of the foils, originally initiated by the studies done at the Helsinki Institute of Physics (HIP) \cite{matti,QAGEM_NIMA}. A high definition optical scanner and an advanced image analysis software, developed at HIP, is capable of measuring parameters of individual holes, such as the diameter of the polyimide hole, and the diameters of the copper holes from both sides. It was shown that gain prediction is possible based on the knowledge of those geometrical hole properties \cite{QAGEM_NIMA}.

The test case back then was based on 10 $\times$ 10\,cm$^2$ double-mask GEM foils. However, only one of the test foils was measured with a position sensitive system with high enough granularity to measure local gain variations. In particular, the foil was placed on top of a Micro-Mesh Gaseous Structure (MICROMEGAS) detector~\cite{micromegas} with x-y strip readout. The active area \mbox{was 89.6\,mm $\times$ 89.6\,mm} covered by 358 strips per dimension. The pitch of the readout strips was 250\,$\mu$m with the y-strips being 80\,$\mu$m wide and the x-strips, located below, 200\,$\mu$m for equal charge sharing\,\cite{Rui}. The gain maps were binned with a bin size of 4.5~$\times$~4.5\,mm$^2$. The published results were encouraging but a refined study was deemed necessary for a quantitative conclusion.

In the follow up study, which is the subject of this paper, we tested three new GEM foils produced with double-mask technique with different setups. However, at first we did not find similar inverse dependency of the hole diameters to the gain than with the foils discussed in our earlier study~\cite{QAGEM_NIMA}. We then looked into combined features that could have an effect on the electric field and thus the gas gain. By combining the optical scanning information from both sides of the GEM foils we found out that there is occasionally a small but measurable offset between the matching copper holes. This offset or misalignment can also be seen in the asymmetry of the polyimide rims indicating that the hourglass shape of the hole is in fact skewed. 

In the following, we describe the new measurements of hole offset that lead us to an improved understanding of the correlation between the gas gain and the geometrical hole properties. In section~\ref{sec_setup} we present the measurement setups in detail, followed by a description of the analysis methods in section~\ref{sec_ana}, and a discussion of the results in section~\ref{sec_results}. Last but not least, we compare the measured results to gain simulations of misaligned GEM holes in section~\ref{simulations}.

\section{Test setup and measurements}\label{sec_setup}

In this study two types of measurements were performed for every GEM foil: one using the high definition optical scanning system in order to take images throughout the whole surface of the copper electrodes from both sides and one using dedicated setups to measure the gain uniformity of the foils under X-ray exposure.

\subsection{Optical Scanning System}

\begin{figure}[]\centering
\includegraphics[width=0.39\textwidth,clip]{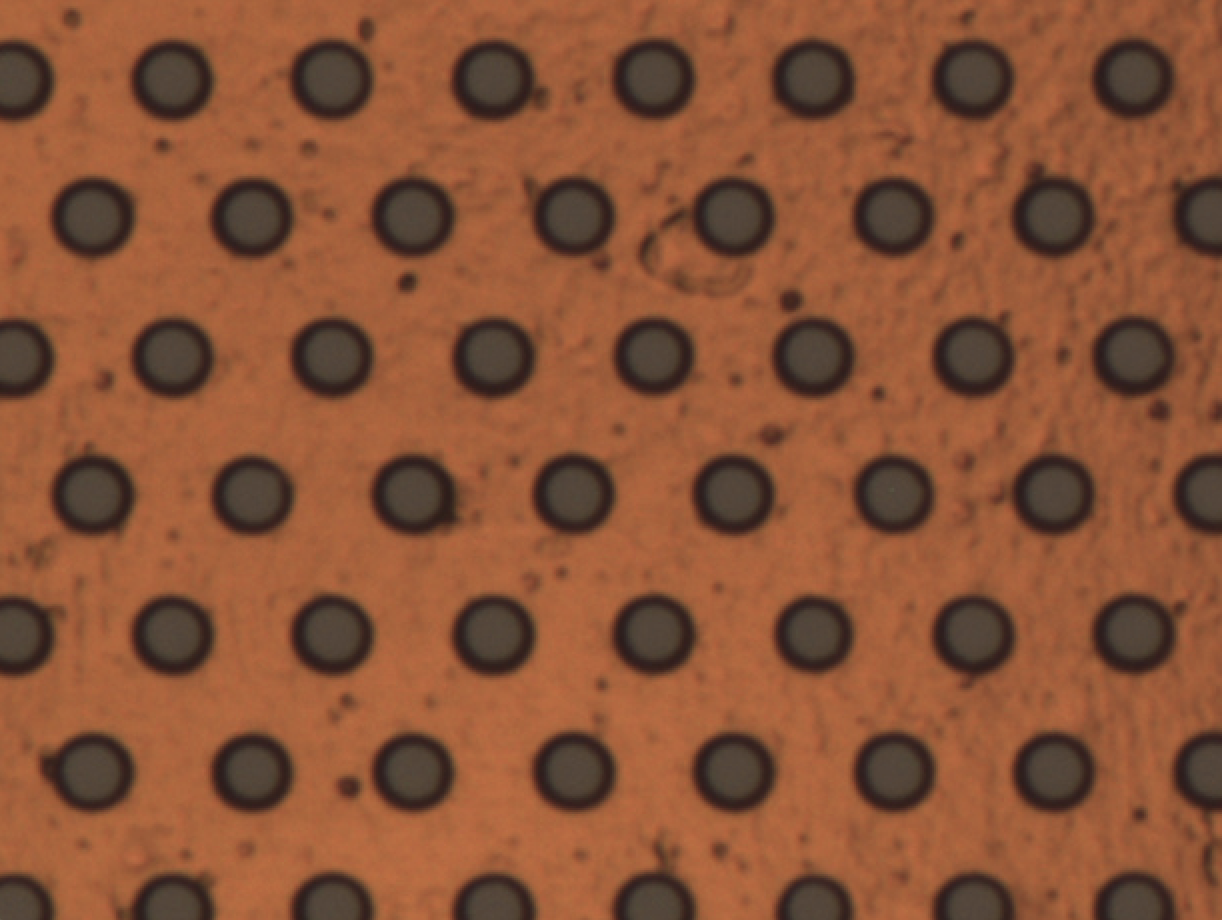}
\caption{Image of an excerpt of a GEM foil taken with the high definition optical scanner.}
\label{fig-ossimg}
\end{figure}

A unique large area Optical Scanning System (OSS) was developed at the Detector Laboratory of Helsinki Institute of Physics and the Department of Physics of the University of Helsinki.  The development was driven by the needs of the Quality Assurance (QA) of the GEM-foils for the SuperFRS experiment at FAIR~\cite{mattivaitos} and the GEM readout chambers of the upgrade for the TPC of the ALICE experiment at CERN~\cite{TPCTDR}. Details about the hardware can be found in reference \cite{matti} and the image reconstruction framework is described in detail in reference \cite{QAGEM_NIMA}. In short, the scanner consists of a robotic x-y-z table with a machine vision camera coupled to a telecentric optics. The camera has an effective pixel size of 1.67\,$\mu$m and the optics has a 1$\times$ magnification. An example of an excerpt of an unprocessed image of a GEM foil is shown in Fig.\,\ref{fig-ossimg}. The analysis software is capable of extracting GEM hole diameters (polyimide, copper) and the hole pitch from such RAW images with a precision close to 0.5\,$\mu$m.  

\subsection{Gain measurement system}
To evaluate the uniformity of the gain distribution and to determine the gain maps of the GEM foils two different setups were used. Both setups include a GEM foil under test (TG) and a reference detector. In the first setup, the reference detector is a double GEM unit while in the second one, it is a Multi Wire Proportional Chamber (MWPC)\,\cite{MWPC_Varga}. The reference detectors are used in order to detect the ionization that takes place above and below the TG. In both setups, a 2D strip readout geometry is used for the reconstruction of the 2D maps. \mbox{An\,$^{55}$Fe} source with the main emission line at 5.9\,keV is used to illuminate the entire GEM foils and to measure their gain.

\subsubsection*{Setup 1}

\begin{figure}\centering
\includegraphics[width=0.45\textwidth,clip]{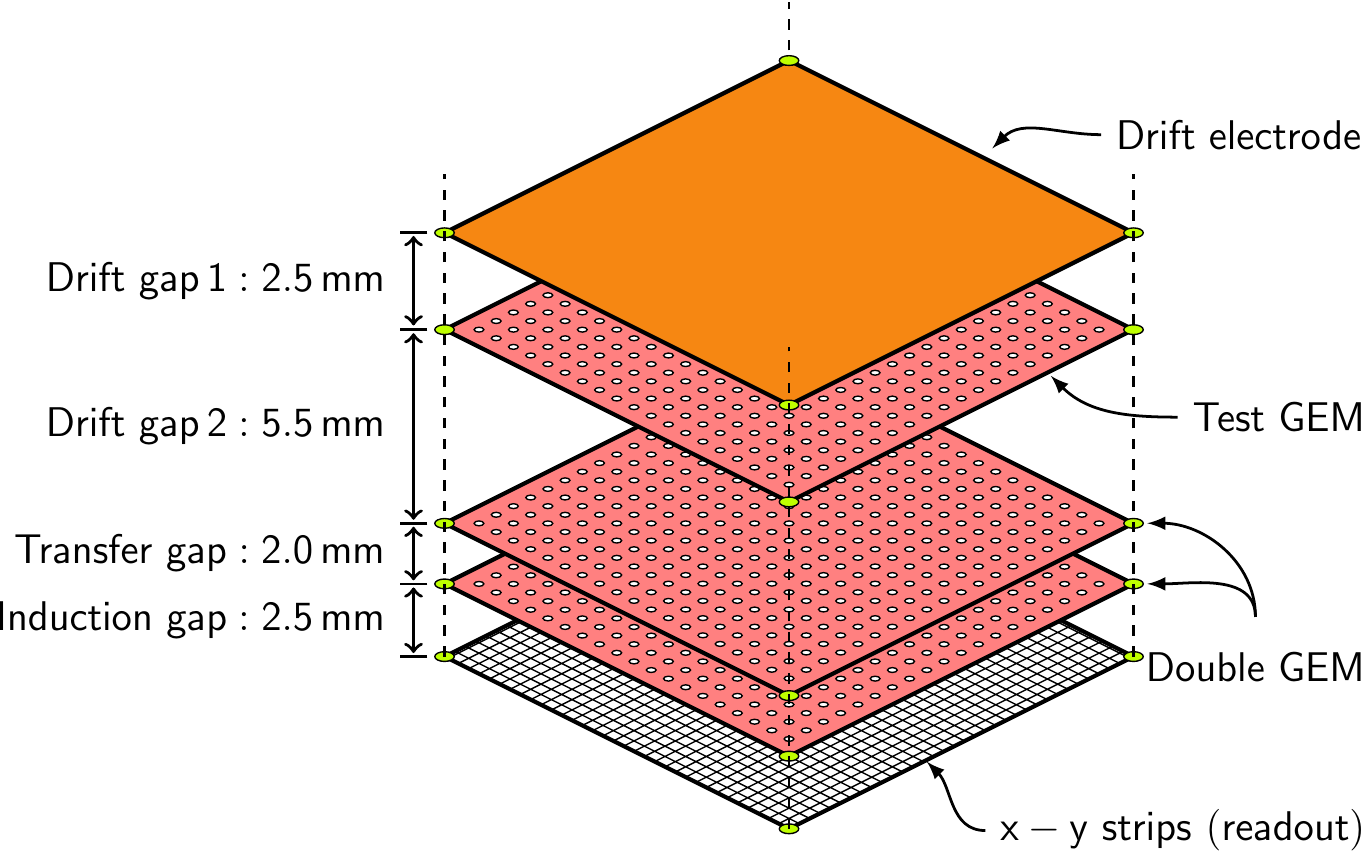}
\caption{Schematic drawing (not to scale) of the setup used with the double GEM as a reference detector for the gain measurements.}
\label{fig-setup}
\end{figure}

In the setup 1\footnote{The measurements using this setup were performed at Helsinki Institute of Physics.}, the GEM foils are introduced between the cathode (drift electrode) and the anode (x–y readout strips) as shown in Fig.\,\ref{fig-setup}. The TG is placed 2.5\,mm below the drift electrode. This gap is defined as Drift gap 1. The double GEM unit is placed 5.5\,mm below the TG (Drift gap 2). The two GEMs, in the so-called double unit, are 2 mm apart and this gap is defined as the Transfer gap. The bottom GEM of the double GEM unit has a distance of 2.5 mm from the readout layer (Induction gap). 

The readout layer consists of 256 $\times$ 256 x-y strips with an active area of 102.4\,mm $\times$ 102.4\,mm. The x- and y-strips have a pitch of 400\,$\mu$m and their width is~350 and 80\,$\mu$m respectively, with the x-strips being below the y-strips\,\cite{Rui}. The different strip widths are important for equal charge sharing between the x- and y-strip layer. 

The whole system is enclosed in a gas tight box, flushed with an Ar-CO\textsubscript{2} gas mixture with a ratio of~70/30. The box has a radiation entrance window made of a 50\,$\mu$m thick polyimide foil, with slightly greater surface area than with the GEM foils.
Voltages were supplied by an ISEG NHQ 246L two channel high voltage power supply\,\cite{iseghv}. One channel was used to apply HV on the TG and the drift cathode, and the other was used to apply HV on the double GEM unit.

The data acquisition was handled by the Scalable Readout System (SRS)\,\cite{srsapv} based on the APV25 Front-End ASIC\,\cite{APV25} that is developed within the CERN RD51 community for MPGD research\,\cite{RD51}. The data acquisition software was based on DATE, the Data Acquisition and Test Environment from the ALICE collaboration\,\cite{date}. The event reconstruction software, containing among others the clustering algorithms, has been further developed from the data quality monitoring framework of ALICE, AMORE\cite{amore}. Since the SRS/APV system does not allow self-triggering, the trigger signal was taken from the bottom side of the lower GEM of the double GEM unit. The signal was processed by an ORTEC pre-amplifier 142IH\,\cite{ORTEC_preamp} and and ORTEC 855 DUAL SPEC AMP\,\cite{ORTEC_specamp} and a \mbox{CAEN 4 CHS} DISCRIMINATOR mod.\,84\,\cite{CAEN_n84}.

\subsubsection*{Setup 2}
\begin{figure}\centering
\includegraphics[width=0.4\textwidth]{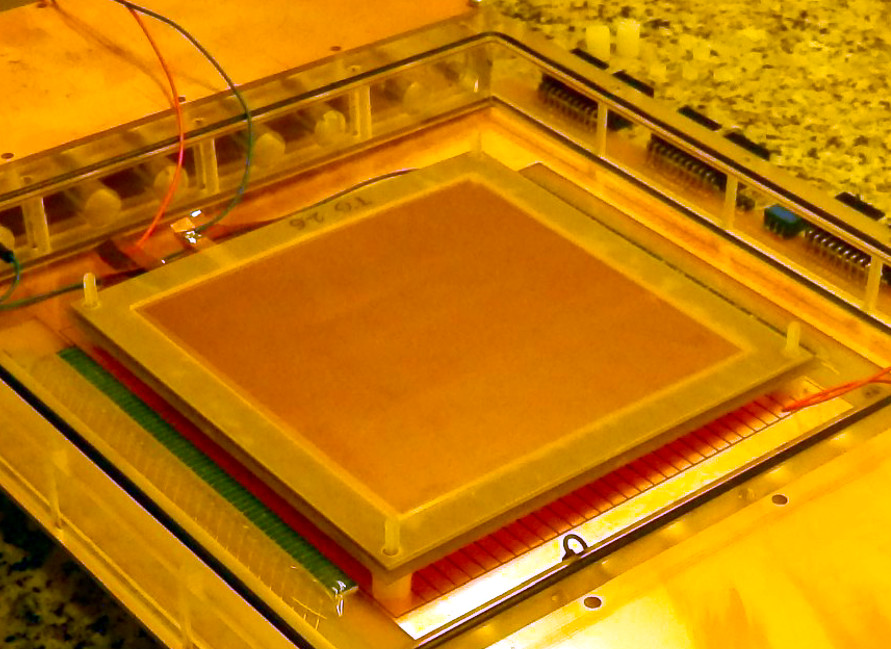}
\caption{Photo of the setup used with the MWPC as a reference detector for the gain measurements. The TG (top layer in this photo) is placed 5\,mm above the MWPC. \label{tg3_mwpc}}
\end{figure}

In the setup 2\footnote{The measurements using this setup were performed at the Wigner Research Center for Physics (RCP) in Budapest.}, the TG is placed on top of a MWPC\footnote{The specific MWPC has been developed by the Wigner RCP, in Budapest\,\cite{MWPC_Varga}. It is a low material budget MWPC, called ”Close Cathode Chamber”.}. This setup is similar to the one used for the QA of the GEM foils for the ALICE TPC upgrade. A short description of the system, shown in Fig.\,\ref{tg3_mwpc}, can be found in Ref.\,\cite{ALICE_QA}.

The MWPC consists of 21\,$\mu$m thin sense wires and~125\,$\mu$m thick field wires arranged in one plane and in an alternate layout. The distance between two sense wires or two field wires is 4 mm. The wires are mounted inclined at a distance of 1.5 – 2.0 mm above the readout electrode. The anode wires for this test were set to~+800 V and the field wires, used for the readout, were set to~-300 V. The readout electrode consists of 28 copper strips, set at ground potential, that are 3.6 mm wide with a pitch of 4\,mm and are placed perpendicular to the wires.  
This allows for a two dimensional readout with a~28 $\times$ 28 pad-like structure.
The data acquisition system is based on a custom made front end card with a 12 bit ADC interfaced to a Raspberry Pi computer. 

We installed the TG at a distance of 5\,mm above the sense and field wires in connection with a guard rim for improving the field uniformity at the detector edges. The drift cathode was placed 5\,mm above the GEM foil. The cathode voltage was set relative to the voltage of the GEM top electrode such that the drift field corresponds to the 0.4\,kV/cm value for all measurements. Also this system is enclosed in a gas tight box flushed with the same Ar-CO\textsubscript{2} mixture~(70:30).

\subsection{Measurements}
As a direct continuation of the study described in Ref.\,\cite{QAGEM_NIMA}, we performed a series of measurements in order to better understand not only the effects of the geometry of the GEM holes to the gain but also the misalignment of the GEM holes between the two electrodes of a foil that is introduced when the double- mask technique is used. The latter can potentially also affect the gain of a detector. For these measurements, we used three new GEM foils called TG1, TG2 and TG3.

First we characterized them with the OSS and produced detailed feature maps, including in particular maps of the copper hole diameters from both sides, and a map of the polyimide hole diameter. Examples of the feature maps from these three GEM foils can be found in the appendix. The process of producing such feature maps from the RAW images of the OSS is described in detail in reference\cite{QAGEM_NIMA} and will not be repeated here. The naming scheme and orientation of the foils are also taken from the paper. The \textit{top-side} refers to the side where the electrode connections are positioned at the lower right corner, and the \textit{bottom-side} where the electrodes are at the lower left side, flipped along the vertical axis (see Fig.\,\ref{GEMorientation}).  

Afterwards, we measured the gas gain uniformity. For the test foils TG1 and TG2 we used the setup 1 with the double GEM unit. We applied a potential difference of around 280~V to the TGs. We set the drift fields DF1 and DF2 to values around 1~kV/cm. Details can be found in Tab.\,\ref{tab_setup}\footnote{Because of changing noise levels and atmospheric pressure changes during the data taking period, we had to slightly adjust voltages on the GEM foils as well as fields in order to have both photo-peaks within the limited dynamical range of the SRS/APV DAQ system.}. For the GEMs in the underlying double GEM unit we applied $\sim$390~V, which results in a transfer field (TF) of $\sim$2450~V/cm, and an induction field (IF) of $\sim$1960~V/cm. 
\begin{table}[]\centering
\caption{Field strengths during measurements for setup 1.\label{tab_setup}}
\begin{small}
\begin{tabular}{lrrr}  \toprule
Run &DF 1 [V/cm]& DF 2 [V/cm]&  TG volt. [V]\\ \midrule 
r306 (TG1 b) & 1011 & 1004 & 275\\
r317 (TG1 t) & 1079 & 782 & 294\\
r323 (TG2 t) & 1018 & 1099 & 277\\
r329 (TG2 b) & 1018 & 1065 & 277\\ \bottomrule
\end{tabular}
\end{small}
\end{table}

\begin{figure}\centering
\includegraphics[width=0.4\textwidth]{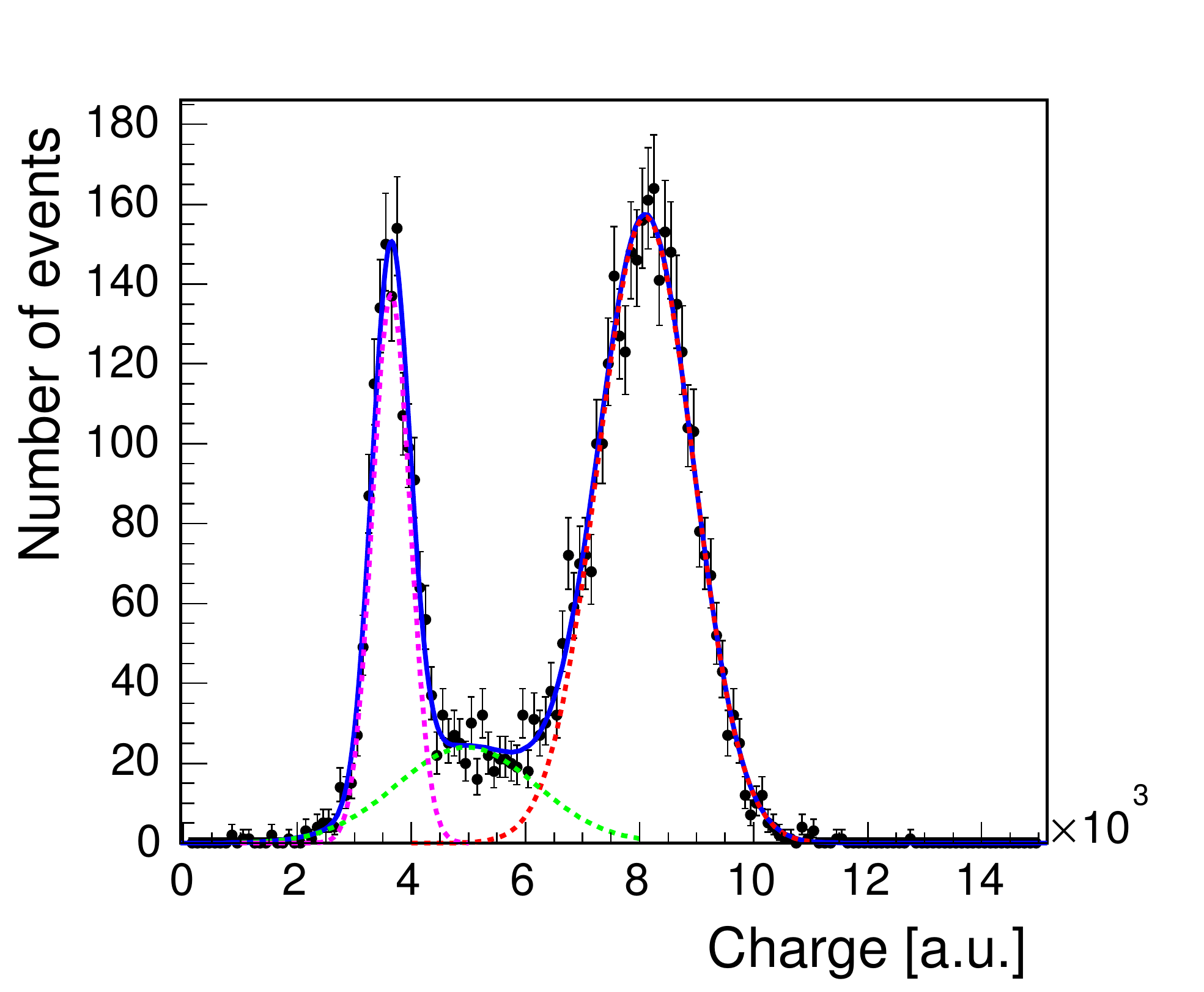}
\caption{Typical charge distribution for events collected in a detector area unit using an $^{55}$Fe source. The plot is taken from Ref.\,\cite{QAGEM_NIMA}. A Micromegas was used instead of the double GEM unit but the principle is the same. Three Gaussian functions were used to fit the data. The peak on the right corresponds to the photo-peak of the GEM plus the Micromegas and the peak on the left to the photo-peak of the Micromegas alone. The fitted curve in the middle accounts for the GEM plus Micromegas escape peak.\label{twinpeak}}
\end{figure}

We used an $^{55}$Fe source to expose the detector to X-ray photons with energies of around 5.9\,keV. From the measurements we can see that we collected two types of events, events where the primary ionization happens between the drift electrode and the TG (DF1 in Fig.\,\ref{fig-setup}), and events where the primary ionization happens between the TG and the double GEM unit (DF2 in Fig.\,\ref{fig-setup}). This can be seen as a double $^{55}$Fe photo-peak in the spectrum (see Fig.\,\ref{twinpeak} that is taken from Ref.\,\cite{QAGEM_NIMA}). 
From the spectrum we can obtain the relative gain of the test GEM by dividing the central value of the photo-peak with higher energy by the central value of photo-peak with lower peak energy. For more details, we refer to our earlier publication\,\cite{QAGEM_NIMA}. In this way we can reduce the effects of the possible non-uniformities in the underlying detector and in the readout electronics. 

To analyse the local relative gas gain variations, we divide the GEM foil area into small segments of equal dimensions by using the 2D strip readout information. Because of the high granularity of the readout strips, we can generate fine structured maps with bin sizes down to 1\,mm$^2$. However, due to the time constraints of our measurement campaign and the limitations in the data handling and storage, we enlarged the bin size to 4\,mm$^2$ approximately. With this bin size we are still sensitive to the local gain variations in the GEM holes. By recording around 5 million events we have enough statistics per segment to ensure an accurate fitting of the spectra and to limit the uncertainties. 

\begin{figure}\centering
\includegraphics[width=0.4\textwidth]{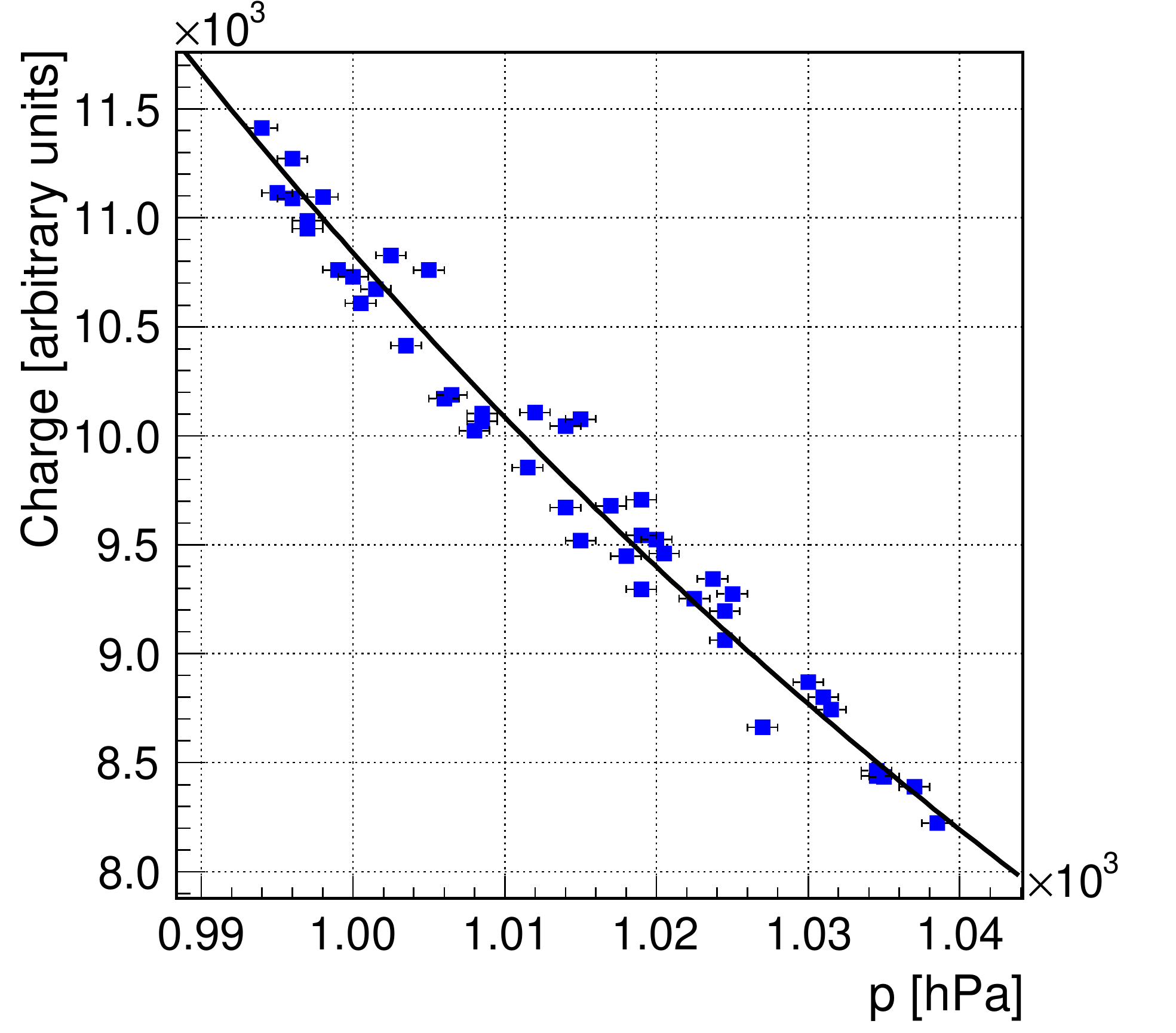}
\caption{Pressure correction function as exponential fit to data.\label{correctionfunction}}
\end{figure}

For every measurement we also measured the running conditions, such as temperature, pressure, voltages and currents. In order to obtain comparable results between the measurements, we used these values to correct the resulting gas gain values. The correction function we used was obtained from a series of measurements with the detector setup 1 using the $^{55}$Fe source. The measurements were made over a period of time, during which the atmospheric pressure changed from~994 up to~1038\,hPa. The voltages and fields were kept constant.  We noticed within the whole measurement campaign that the temperature variation in our lab was within one degree in Celsius. Therefore, the effects of the temperature variations were not used as a parameter in the correction function. The resulting exponential curve fitted to the data, serving as our correction function, is shown in Fig.\,\ref{correctionfunction}. Normalized to the standard atmospheric pressure of 1013\,hPa we get as correction function,
\[\frac{g_{\mathrm{measured}}}{g_{\mathrm{corrected}}}=A\,\exp{\left({\frac{B}{p}}\right)}\]
where $g$ stands for the relative gas gain, $p$ for the atmospheric pressure during data taking. $A$ and $B$ are constants retrieved from the fit to the data (see Fig.\,\ref{correctionfunction}) with values of $(7.60 \pm 0.58) \times 10^{-4}$ and $(72.76 \pm 0.78) \times 10^{2}$\,hPa respectively.
However, it is worth to mention that atmospheric pressure variations act both on the gain of the TG and the underlying double GEM unit. As we extract the relative gas gain of the TG by dividing the photo-peak centroid of the full detector by the photo-peak centroid of the underlying detector we have only a negligible dependency on the atmospheric pressure changes.

Since we have variations in the environmental conditions, such as in the atmospheric pressure, we selected the voltages of the measurement system in such way that we have both photo peaks within well measurable limits of the dynamical range of our readout system. Thus there is a difference in the applied voltages between the measurements.

The GEM TG3 was measured with the setup 2. To obtain similar gain levels as with the setup 1 and in order to compensate for the different environmental conditions between the two laboratories, we applied a potential difference of 300\,V over the foil. However, we also made measurements at a higher potential difference of 400\,V, to study possible changes in the gain uniformity. We set the collection field between the drift cathode and the GEM to 0.4\,kV/cm, and the transfer field between the GEM and the wire plane to 1.4\,kV/cm. For the reconstruction of the relative gain we used the same methods as for setup 1. In addition, we tested various field configurations, and confirmed two earlier findings~\cite{gem_varga}. One is that the relative gas gain variations of the GEM did not substantially change when changing the drift field between 0.2 and 1.2\,kV (full collection). Furthermore, in terms of the transfer field, we found out that the gas gain variations were proportional to the transfer field values in the region between 0.6 and 2.0\,kV.

\section{Analysis}\label{sec_ana}
The data from the optical scanning measurements for all GEMs was reconstructed as described in detail in~\cite{QAGEM_NIMA}.  First, we aligned the data (hole by hole information) to the coordinate system of the gain measurement by rotation and translation.

Next, we produced the relative gas gain maps via custom made software based on the ROOT framework \cite{ROOT}, with automated peak finding and fitting. The two main photo peaks from the ionization regions above and below the TG were fitted. We did not take into account the escape peaks that partly overlap with the photo peaks. Due to the limited dynamical range of the data acquisition system we were not able to adjust the voltages to fully separate the photo peaks from the escape peaks. However, we already estimated in our earlier study that this procedure will introduce an uncertainty value of around 1\% on the measured mean value of the relative gas gain of the TG\,\cite{QAGEM_NIMA}.

\begin{figure*}[ht]\centering
\includegraphics[width=0.315\textwidth]{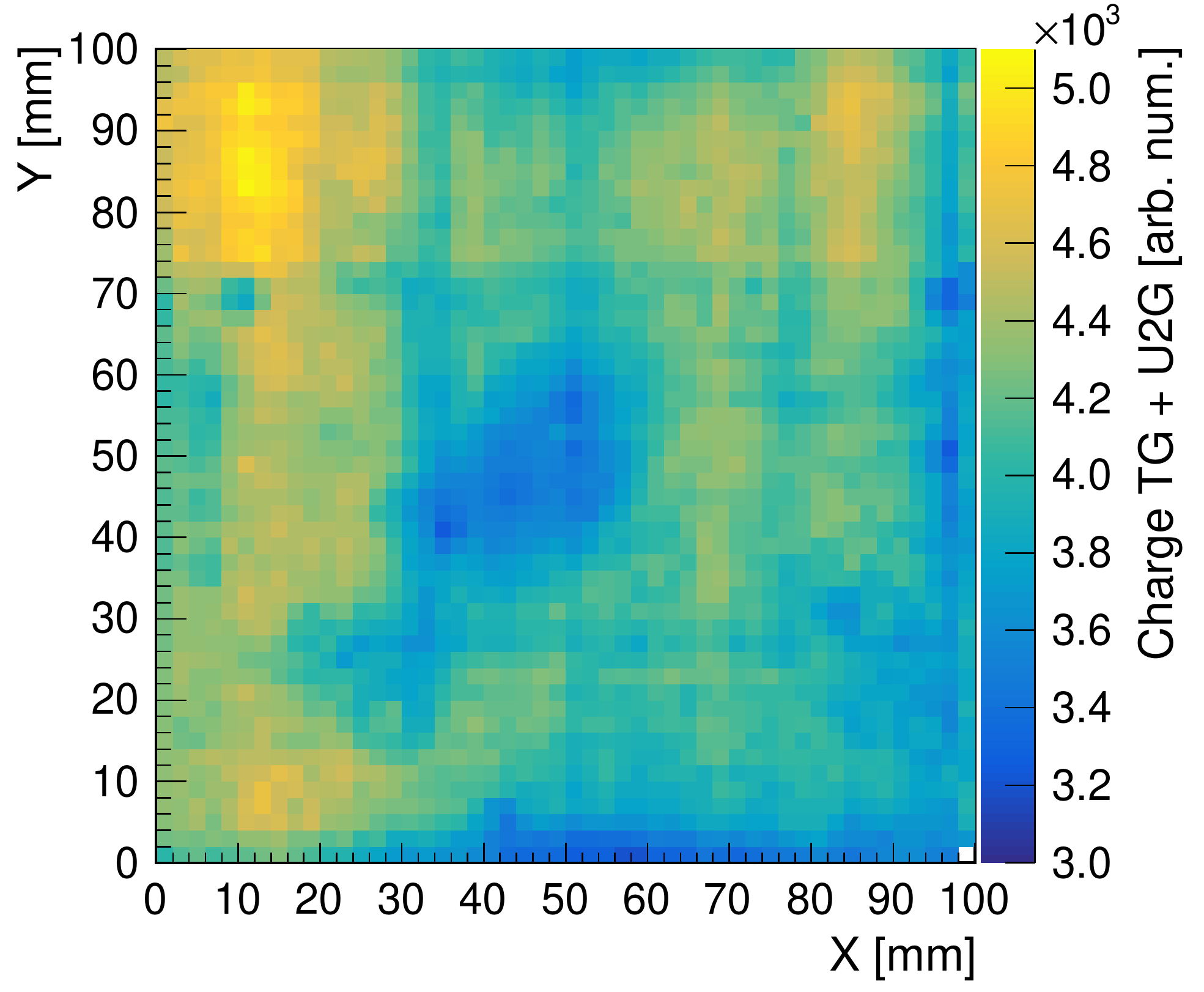}\hspace{1em}
\includegraphics[width=0.315\textwidth]{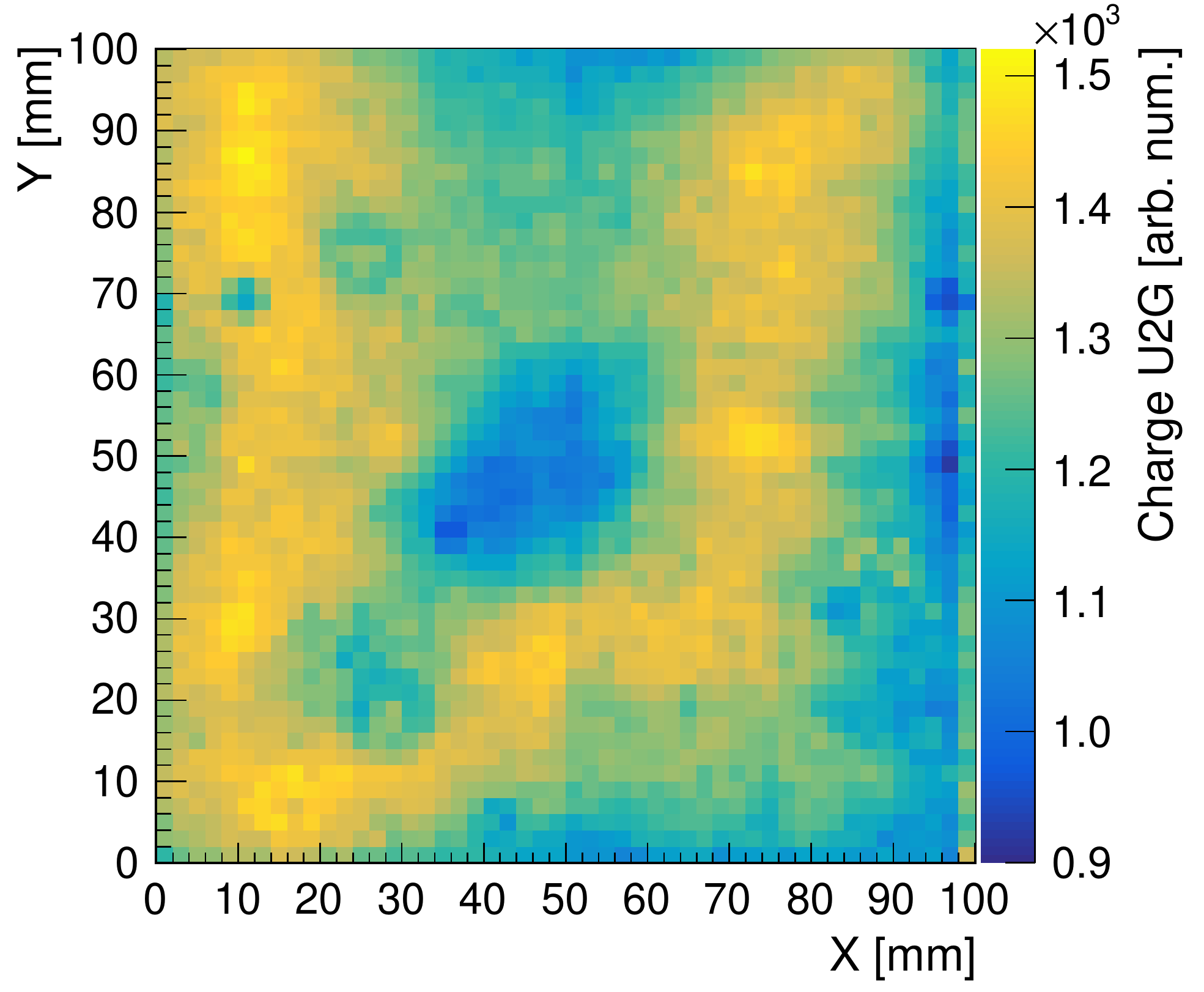}\hspace{1em}
\includegraphics[width=0.315\textwidth]{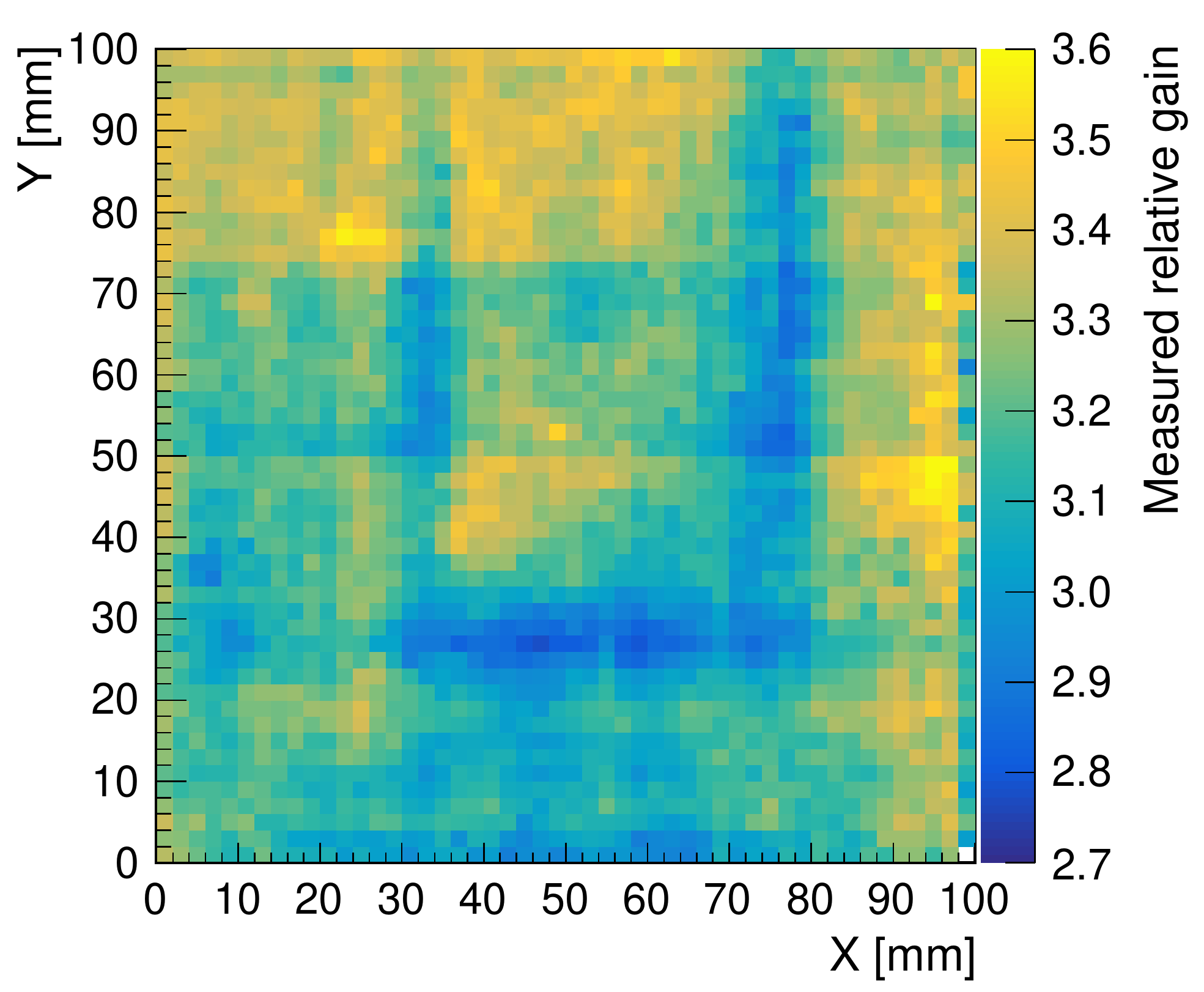}
\caption{Two-dimensional cluster charge maps of the GEM TG1 plus the double GEM unit (left) and of the double GEM unit (middle); Two-dimensional relative gain map (right) obtained by dividing the two-dimensional histogram on the left with the one from the middle.\label{gainmap_measurements}}
\end{figure*}

The procedure is visualized in Fig.\,\ref{gainmap_measurements} for the GEM TG1. A map of the reconstructed photo peak centroids from the ionization above the TG is shown in Fig.\,\ref{gainmap_measurements} (left), and a map of the peak centroids of the ionization below the TG is shown in Fig.\,\ref{gainmap_measurements} (middle). By dividing the values from above and below the test GEM, a relative gain map of the TG is obtained, as shown in Fig.\ref{gainmap_measurements} (right).

To study the correlation between the geometries of the GEM holes and the gas gain, we followed two approaches. First, we undertook a simple bin by bin multiplication of the inverse values of the normalized\footnote{Normalized to the mean value.} features maps, that we then compared to the measured relative gas gain maps. In addition we utilized an artificial neural network (ANN) in regression mode, which we trained with the binned data of the feature maps and the relative gas gain maps of the two GEMs (TG1, TG2), that we measured at HIP. We then used the trained ANN to predict the relative gas gain of the third GEM measured at RCP (TG3). For the ANN we chose the feed forward network (Multi-Layer Perceptron) available from the ROOT framework. As input layer we fed in four features, namely the copper hole diameters from both sides of the GEM foils, the offset values between the top and bottom copper holes and the inner polyimide hole diameter. As output layer we assign the measured relative gas gain values. We used 3 hidden layers with~8,~8 and 2 neurons. The ANN was then trained with 250 epochs using half of the available data of the GEM foils TG1 and TG2 and the other half for validation. We then used the resulting output function to predict the relative normalized gas gain values for the GEM foil TG3, that was not used in the training and in addition measured with the completely independent gain measurement system, described here as setup 2.

\section{Results}\label{sec_results}
\begin{figure*}[!ht]\centering
\includegraphics[width=0.315\textwidth]{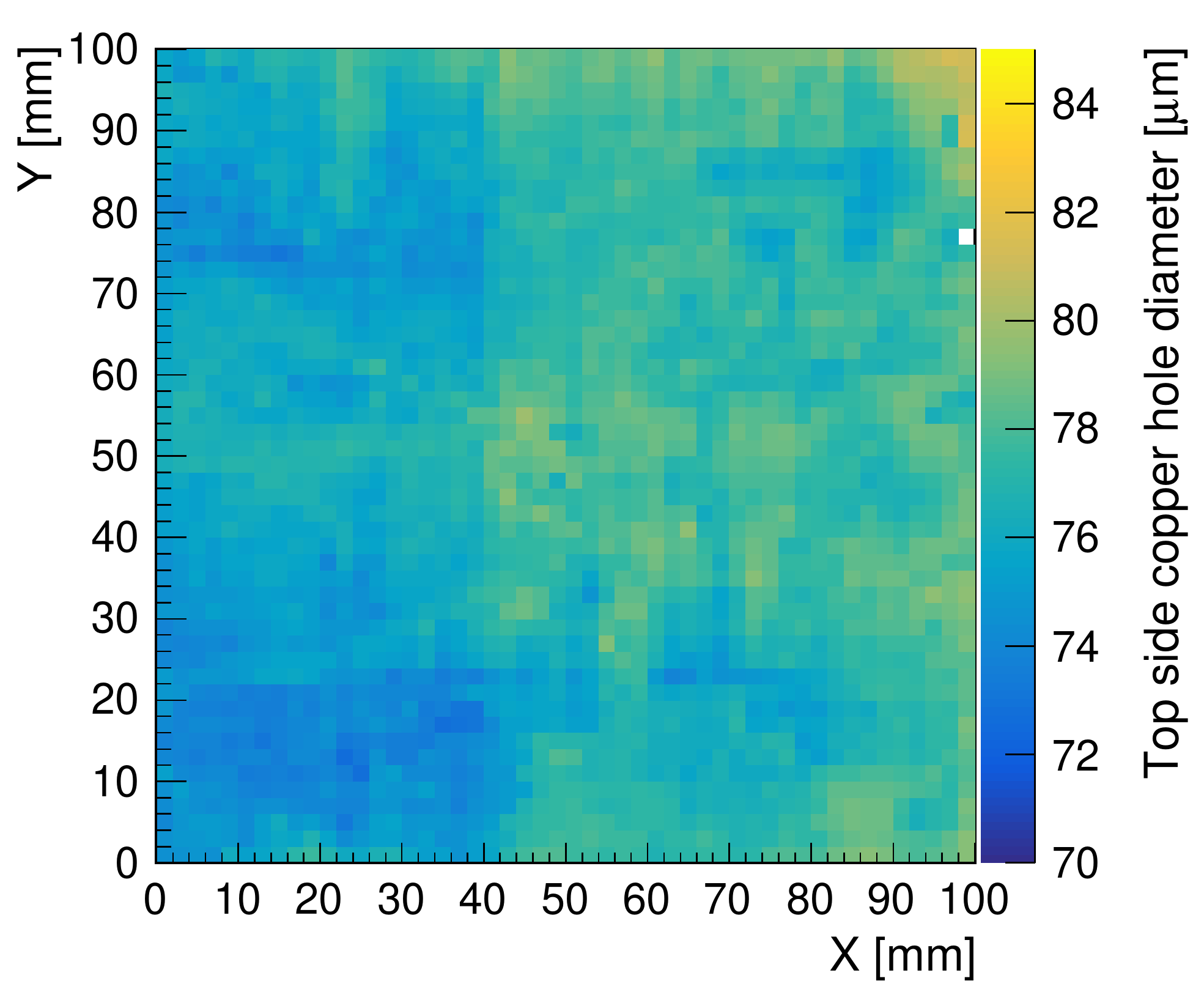}\hspace{1em}
\includegraphics[width=0.315\textwidth]{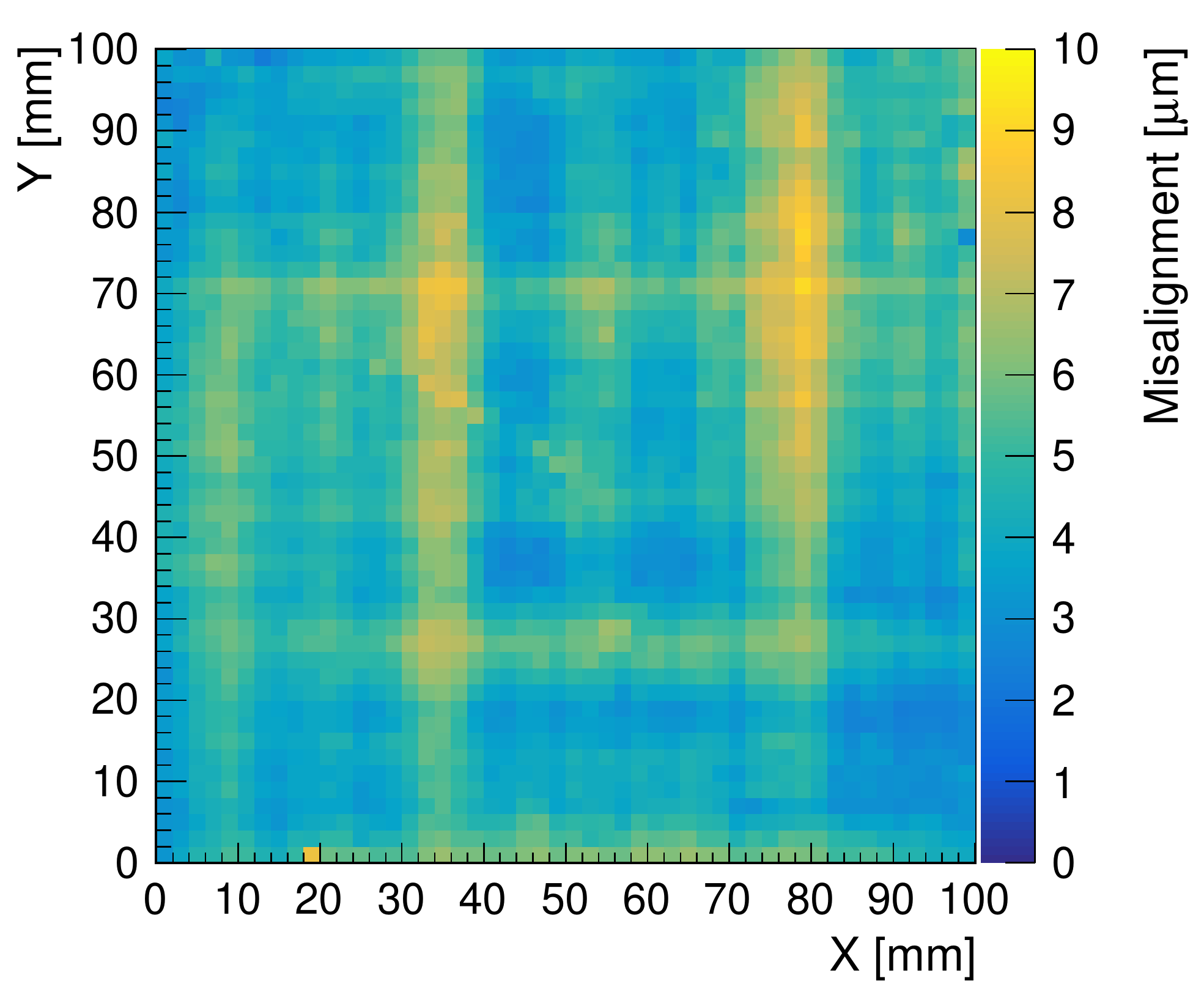}\hspace{1em}
\includegraphics[width=0.315\textwidth]{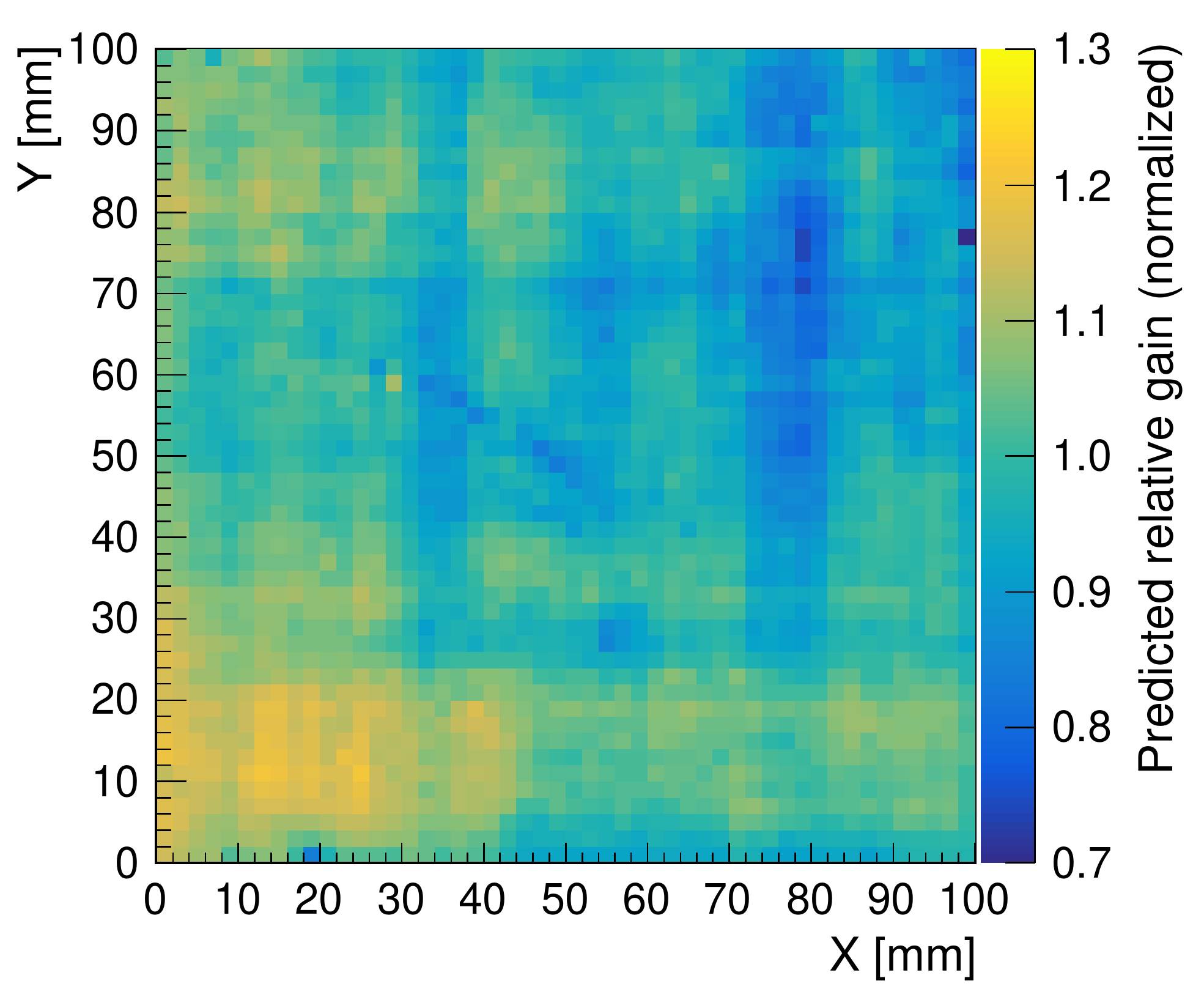}
\caption{Feature maps of the TG1 foil; Two-dimensional histograms of the measured diameters of the outer copper holes on the top side of the foil (left) and the measured misalignment between the top and the bottom copper holes (middle). The two-dimensional histogram on the right is the normalized predicted gain obtained by multiplying the inverse values of the two-dimensional histograms of the top and bottom copper and inner polyimide hole diameters with the two-dimensional histogram in the middle. Bin width is 4 mm$^2$. Measurement values are in $\mu$m.\label{TG1features}}
\end{figure*}

In our earlier study, a direct inverse dependency with the hole diameter was clearly visible. However, in the current measurements with TG1, one can detect only some local areas that show this inverse dependency of the hole sizes with the gas gain values. For the rest of the foil, no dependency can be observed, as can be seen by comparing the copper hole diameter map, shown in Fig.\,\ref{TG1features} (left), with the relative gas gain map, shown in Fig.\,\ref{gainmap_measurements} (right). 
\begin{figure}[]\centering
\includegraphics[width=0.4\textwidth]{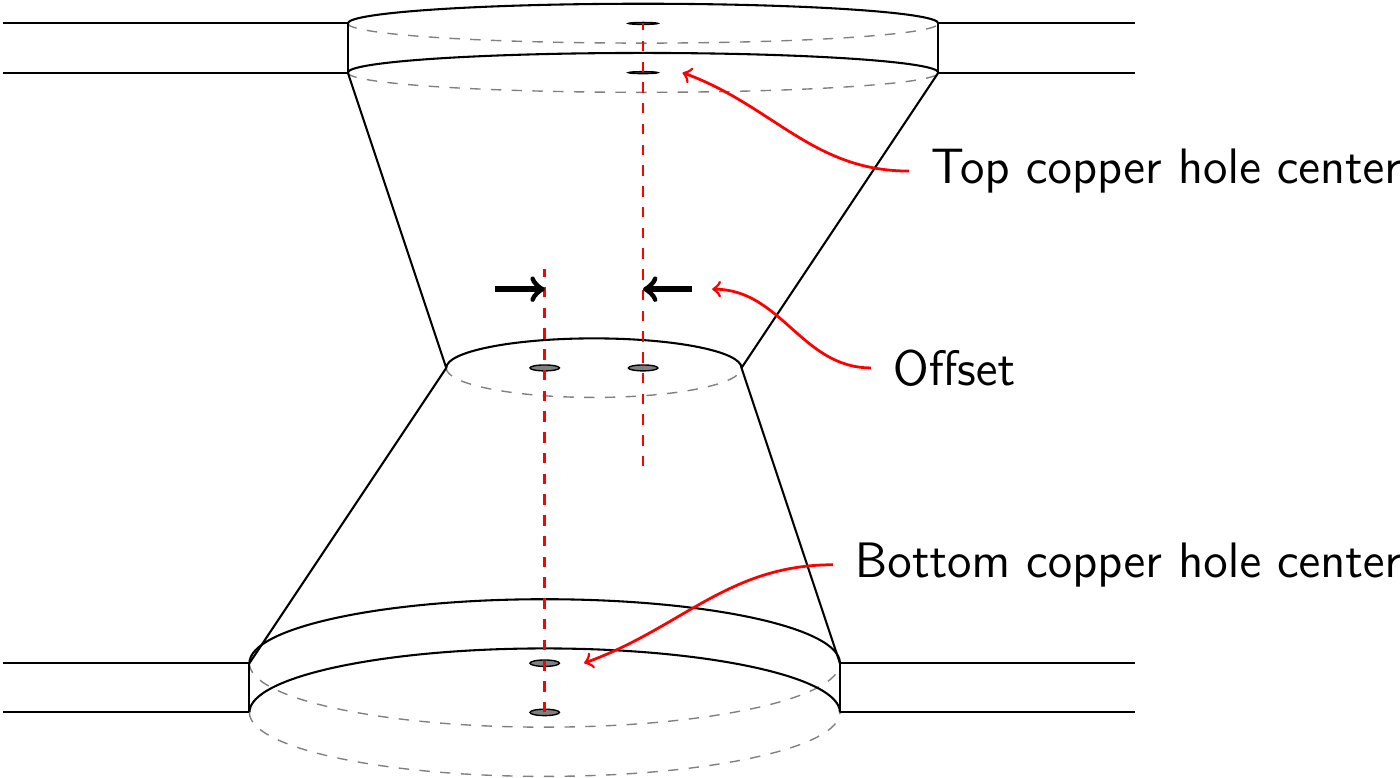}
\caption{Schematic drawing (not to scale) showing the offset between the top and bottom hole introduced by the misalignment of the two photo-masks in a GEM foil when the double-mask technique is used. \label{offset_definition} }
\end{figure}
However, by comparing the absolute coordinates of the centroids from the top and the corresponding bottom copper hole, a measurable offset between the two sides was found as illustrated in Fig.\,\ref{offset_definition}.

\begin{figure*}[!hb]\centering
\includegraphics[width=0.4\textwidth]{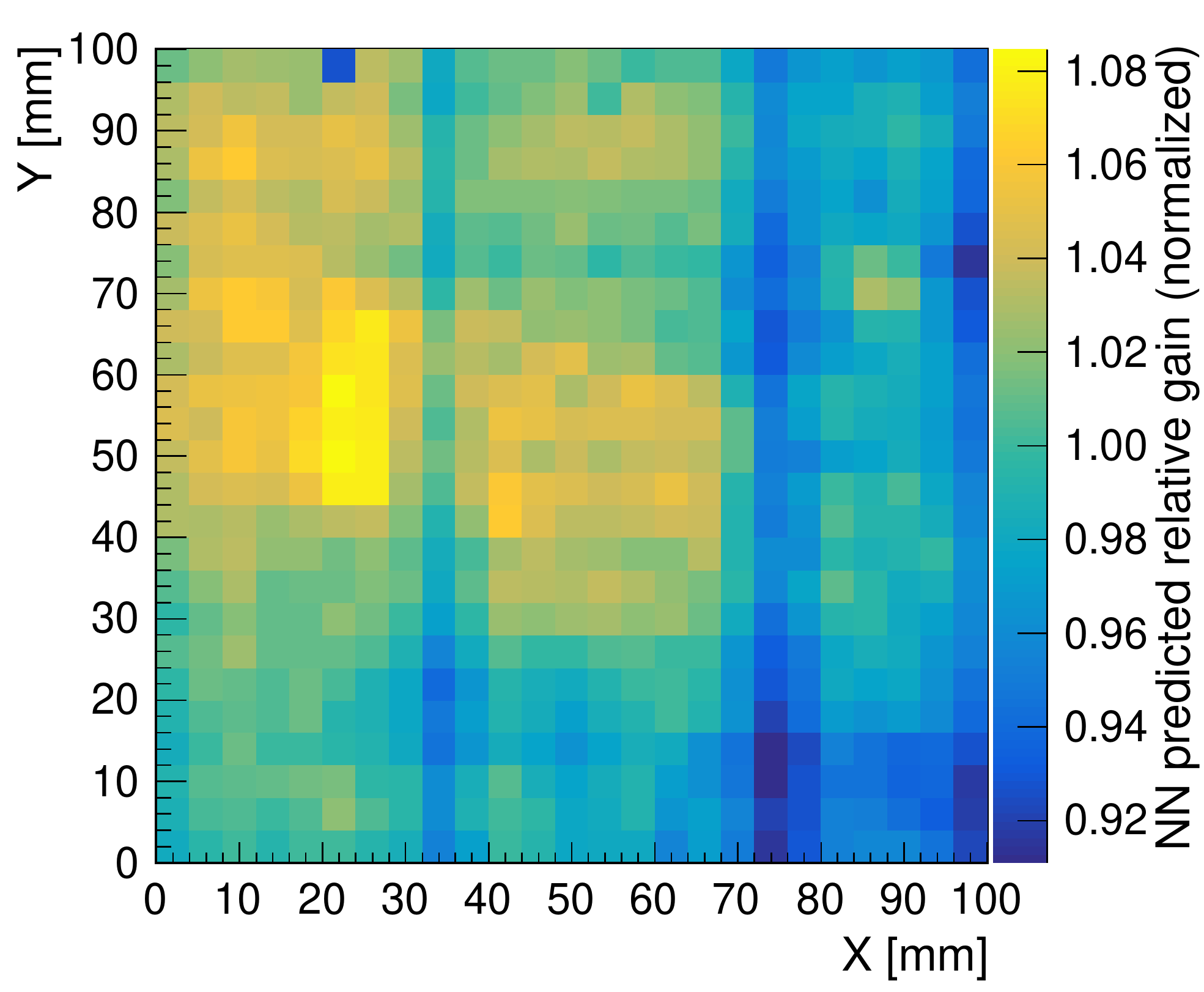}\hspace{6em}
\includegraphics[width=0.4\textwidth]{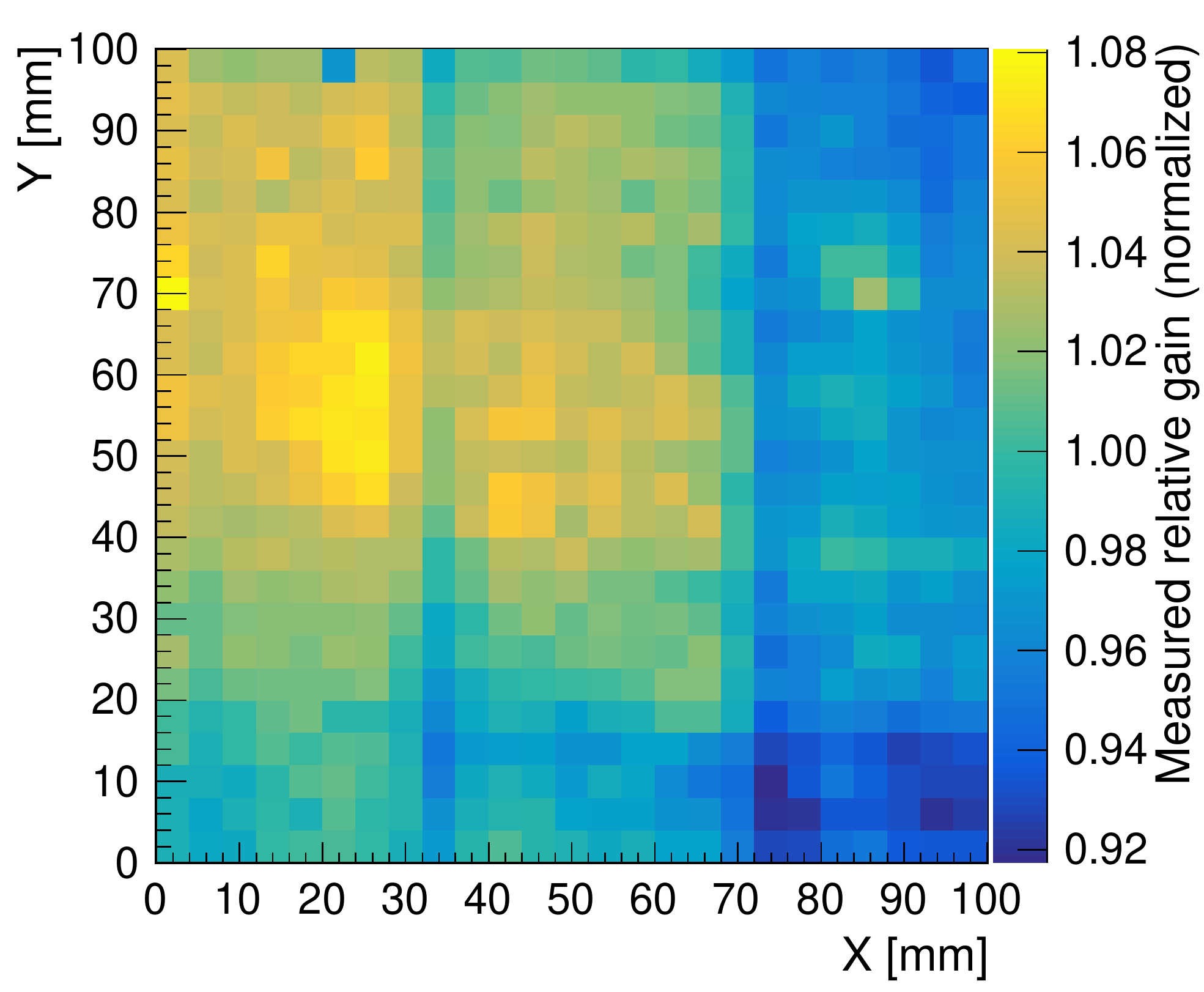}
\caption{(left) Two-dimensional gain map as predicted with the artificial neural network for the TG3 foil; (right) Two-dimensional measured relative gain of the TG3 foil.\label{tg3_comp}}
\end{figure*}
We then produced precise offset maps. Measuring the misalignment of the GEM holes directly from the coordinate data would have necessitated stitching the images with micrometer accuracy over the whole 10 by 10 cm foil, which would have been quite a challenge. Instead, we measured the offset between the centroids of the outer and the inner holes.
To confirm that the holes have indeed a skewed shape as depicted in Fig.\,\ref{offset_definition}, we reconstructed the orientation of the offset for each side of the GEM foil and compared them. The GEM foils were scanned first with the top side facing up. The bottom side was then scanned by flipping the foil around the y-axis. This means that skewed holes will not show any change of the sign for the x-component but the sign of the y- component will change. This is exactly was what observed and the holes that show a measurable offset are indeed skewed.

A map of the misalignment values for the TG1 foil is shown in Fig.\,\ref{TG1features} (middle), and it is clearly visible that the structures seen in the map coincide with the areas with a lower gain in Fig.\ref{gainmap_measurements} (right). The misalignment shows an inverse dependency, similar to the hole diameters.  We now used the inverse of the misalignment map, that we normalized to the mean of the misalignment distribution. This we multiplied to the inverse of the hole diameter map. The result can be seen in Fig.\,\ref{TG1features} (right) and shows similar structures as in the gain map.

\begin{figure}[]\centering
\includegraphics[width=0.45\textwidth]{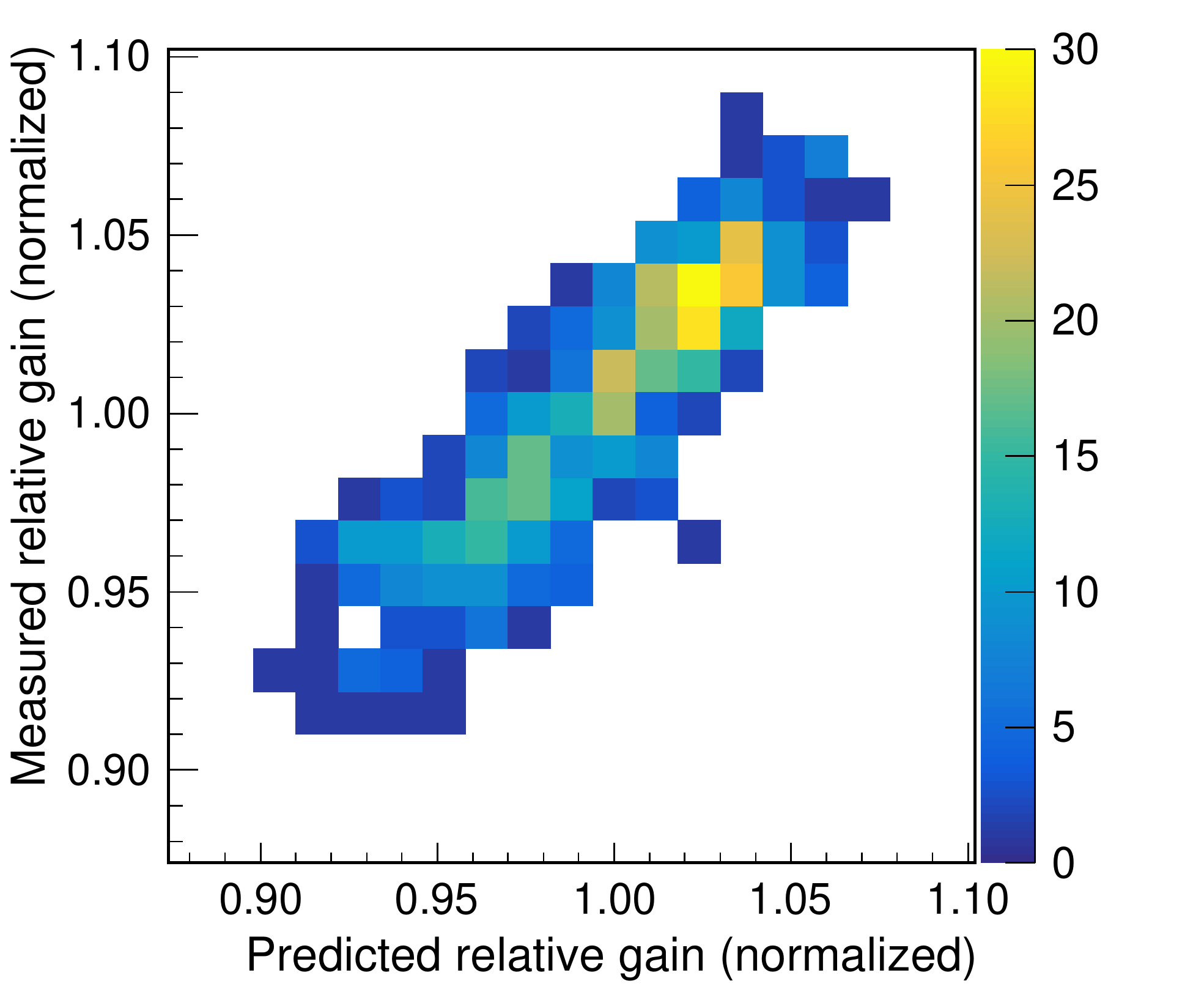}
\caption{Measured relative gain as a function of the gain predicted with the artificial neural network for the TG3 foil.\label{tg3_corr}}
\end{figure}

\begin{figure}[]\centering
\includegraphics[width=0.45\textwidth]{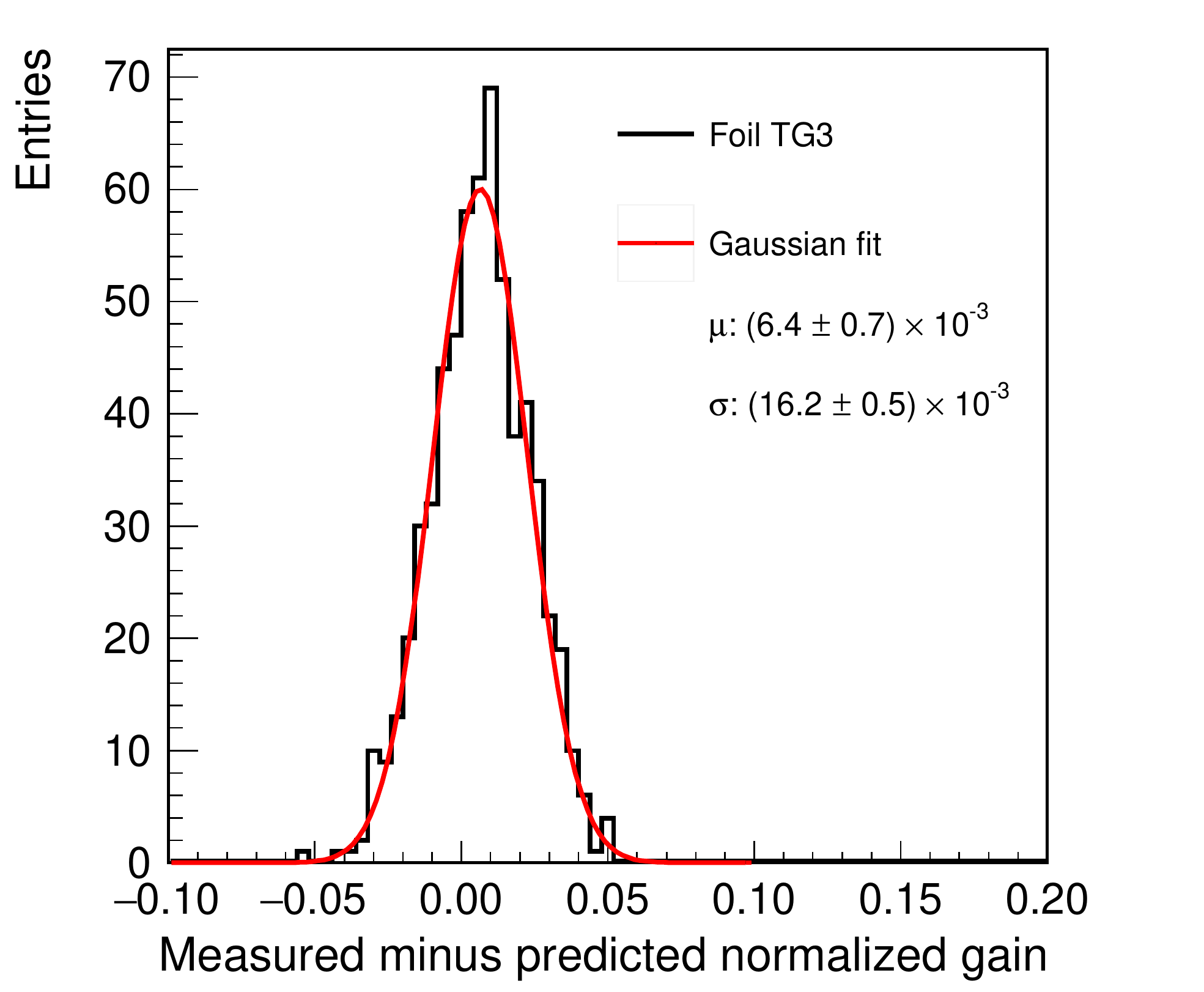}
\caption{Difference between the predicted and the measured normalized relative gain for the TG3 foil (the difference has been obtained by a bin-by-bin subtraction of the two quantities).\label{tg3_err}}
\end{figure}

The results of applying the ANN to the parameters can be seen in Fig.\,\ref{tg3_comp} as direct visual comparison of the predicted gas gain map (left) for the GEM TG3 with the measured one (right). 
Both maps were normalized to the mean values of the relevant gas gain distributions. The correlation plot of the predicted versus the measured normalized gas gain is shown in Fig.\,\ref{tg3_corr} with a correlation factor of~0.88 and a covariance value of~0.0011. The histogram of the bin-wise subtraction of the measured and the predicted gain is shown in Fig.\,\ref{tg3_err}. The Gaussian fit shows a mean of $0.0064\pm0.0007$ and a sigma of $0.0162\pm0.0005$.  
It is interesting to see that the foils that were used for training the ANN show a rather strong misalignment. This misalignment feature receives a strong weight in the trained network.  The TG3 foil on the other hand shows much lower misalignment values than the training foils, with a smooth gradient along the x-axis (see Fig.\,\ref{smooth}). Despite that we see a good relative gas gain prediction power of the trained ANN for the TG3 foil, which suggests that the ANN is trained well by making use of all available features. 

Studying the feature maps of GEMs (shown in \ref{tgfeatures}) it is evident that the offset between the top and the bottom outer copper holes plays a role in the gas gain performance of the GEM. 
To measure the dependency of the gas gain to the hole misalignment, we extracted a subset from our binned data containing holes with diameters close ($\pm 1\,\mu$m) to the typical design values with an outer/copper hole diameter of 70\,$\mu$m on both sides and an inner/polyimide hole diameter of 50$\mu$m.

\begin{figure}[]\centering
\includegraphics[width=0.45\textwidth]{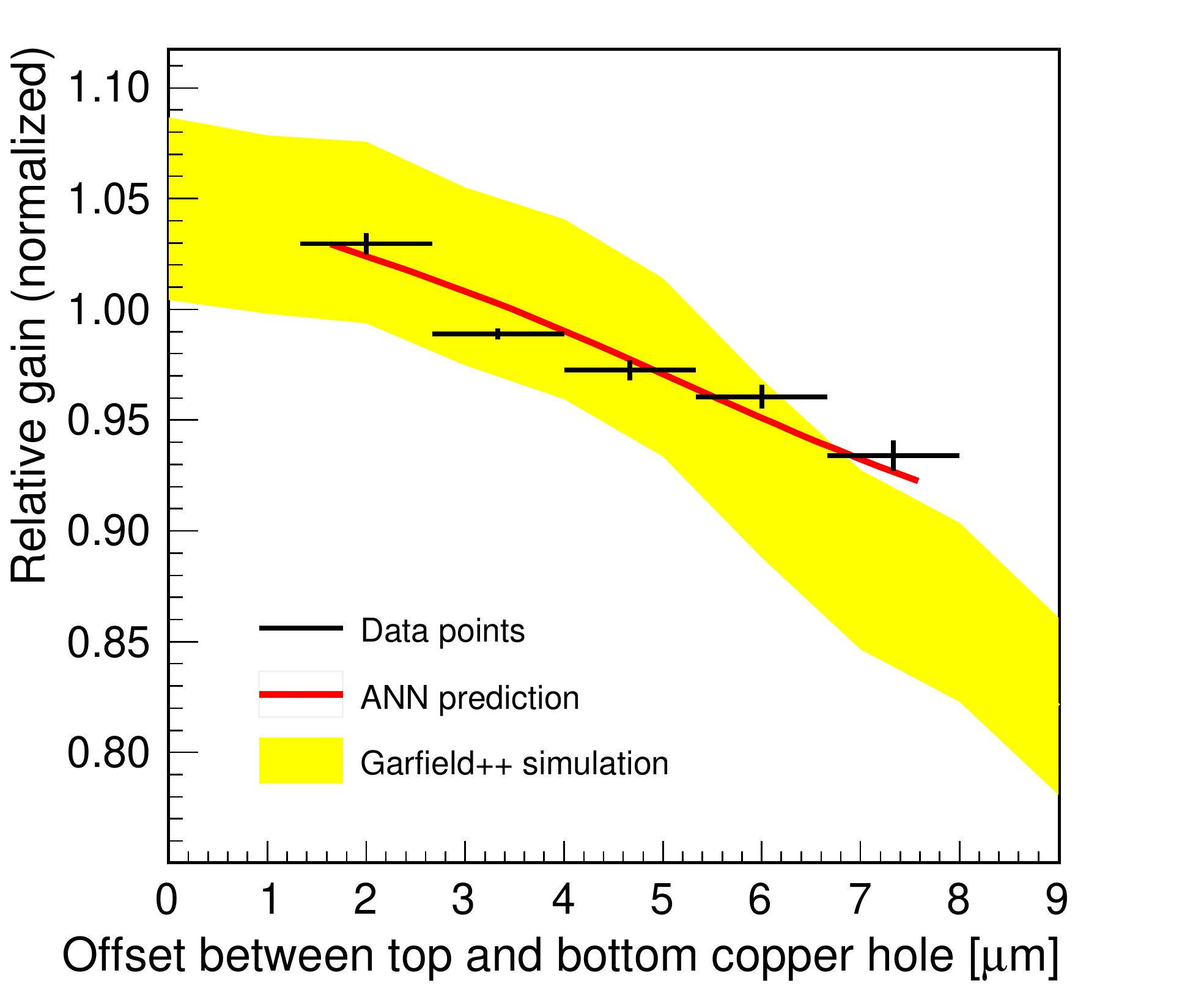}
\caption{Dependence of the relative gain to the misalignment of the top and bottom copper holes for the standard hole geometry of 70-50-70\,$\mu$m. The red curve represents the predictive results from the trained artificial neural network. The yellow band represents the one sigma uncertainty region of the simulated gas gain using Garfield++.  \label{misalignmentgraphs}}
\end{figure}

The result is shown in Fig.\,\ref{misalignmentgraphs}. The red curve in the graph is the predicted gain variation due to the offset by the trained ANN. The curve is consistent with the measured data points and it suggests that the offset is well described by the network. One should note that the relative gain is normalized to the measured gain distribution of the entire foil. This means that the relative gain of a GEM hole is not necessarily at a value of one. 

We can thus conclude that we have a relative gain drop of  $\sim$2\,\% per 1\,$\mu m$ offset.

In addition, we would like to report that the GEM TG3 measured at a higher potential difference of 400\,V shows similar gas gain maps but with a slightly increased non-uniformity. Comparing it with the 300\,V, a simple linear power law can be extracted.

\section{Simulations}\label{simulations}

\begin{figure}[]\centering
\includegraphics[width=0.45\textwidth]{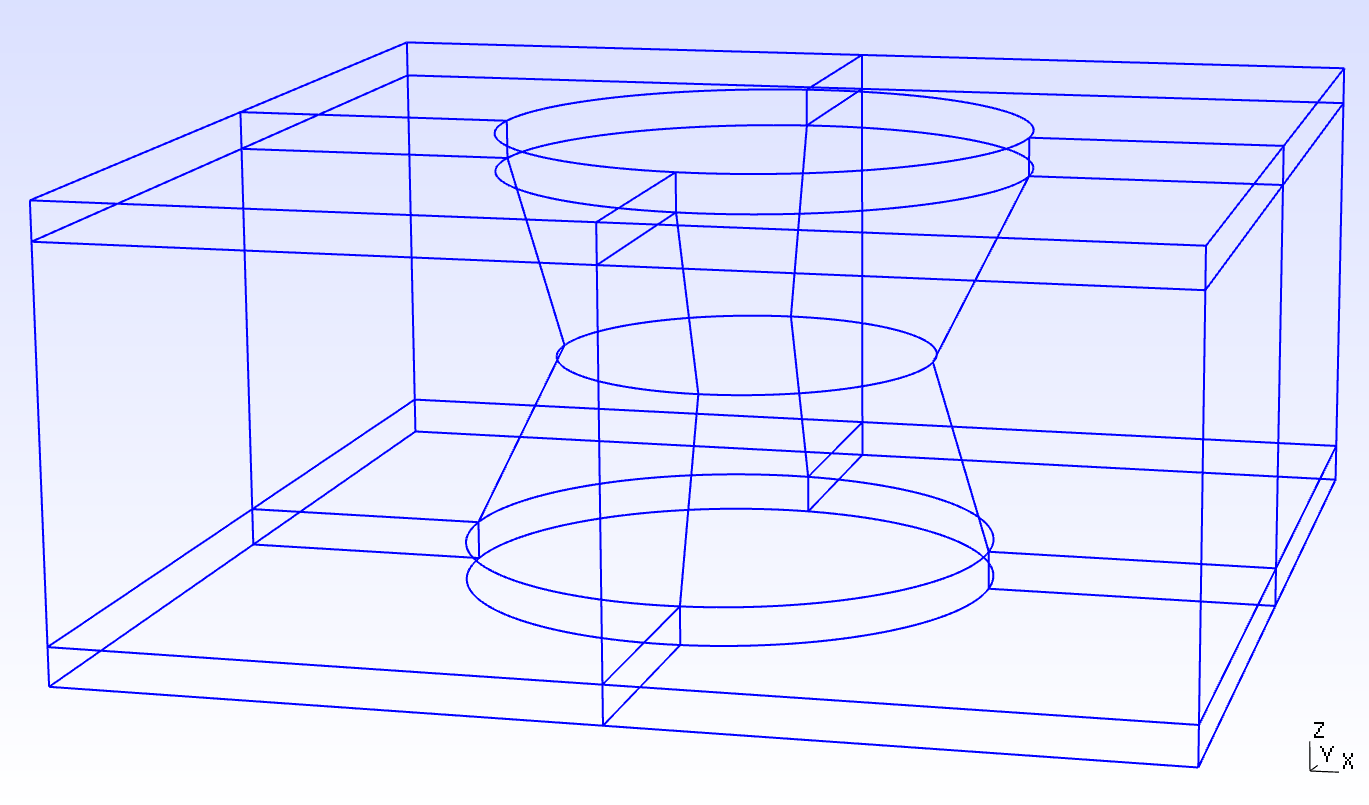}
\caption{Example of an unitary cell used for the simulation. This cell shows a GEM hole with an misalignment of 5\,$\mu$m between the upper and lower copper hole centroid. \label{GEMcell}}
\end{figure}

The effect of the misalignment of the holes to the detector gain was simulated using the \texttt{Garfield++} tool-kit \cite{Veenhof,garfieldpp}. The field-maps needed as input to \texttt{Garfield++}, were generated using the open source finite element mesh generator \texttt{Gmsh} \cite{gmsh}.
We defined the geometries of a minimal GEM element with varying misalignment from 0 to 9\,$\mu$m (for illustration see Fig.\,\ref{offset_definition}). Our unitary cell was a full GEM hole as shown in Fig.\,\ref{GEMcell}\footnote{We are aware of that such a unitary cell can not represent a standard GEM foils with a hexagonal hole pattern. However, we are solely interested in the relative gain change due to misalignment and not in the absolute gain performance of such a GEM foil.}. 
The generated meshes were then processed with the open source multi-physical simulation software \texttt{Elmer} \cite{elmer} to compute the electrostatic field maps. For the simulation we set the gas and fields close to our measurement conditions, with a drift field of~800\,V/cm, the transfer field of~1000\,V/cm and a potential difference of~300\,V for the single GEM foil. Electrons were randomly injected, using a uniform distribution over a single GEM hole unit cell in a distance of~150\,$\mu$m above the center. The electrons that reached the gas volume of~100\,$\mu$m below the center of the GEM foil were used to estimate the relative gas gain. 

We noticed a dependency of the simulated gain to the quality of the generated meshes. As the gain is expected to be identical independent of the orientation of the holes, we tuned the meshes such that the results for the left and the right misalignment showed a similar gain. We then repeated the simulations ten times to estimate the statistical spread. A systematic uncertainty was estimated running the simulation with meshes of varying quality. From the resulting distribution we used the standard deviation and added it to the statistical uncertainty in quadrature.  The result can be seen in Fig.\,\ref{misalignmentgraphs} as yellow band which consist of the simulated relative gain value plus a one sigma uncertainty region. For comparison, we normalized the simulated relative gain values to data. This is justified as we are only interested in the relative gain changes and not on absolute values of the simulated gain, which have a still not fully understood offset to measured gain values especially for argon gas mixtures (see discussion for example in~\cite{mm_avalanche_sim}). More important is the slope of the gain variation compared to the offset of the copper holes. The slope of the simulated curve is reasonably consistent with the measured curve. One should note that the model of the GEM holes for the simulation is only an approximation to the real skewed GEM holes. The inner polyimide hole is not an ideal circle but it has rather an elliptic shape, and the plane of the hole is not necessarily in a plane parallel to the copper plane.  

\begin{figure}[]\centering
\vspace{-0.9em}\includegraphics[width=0.48\textwidth]{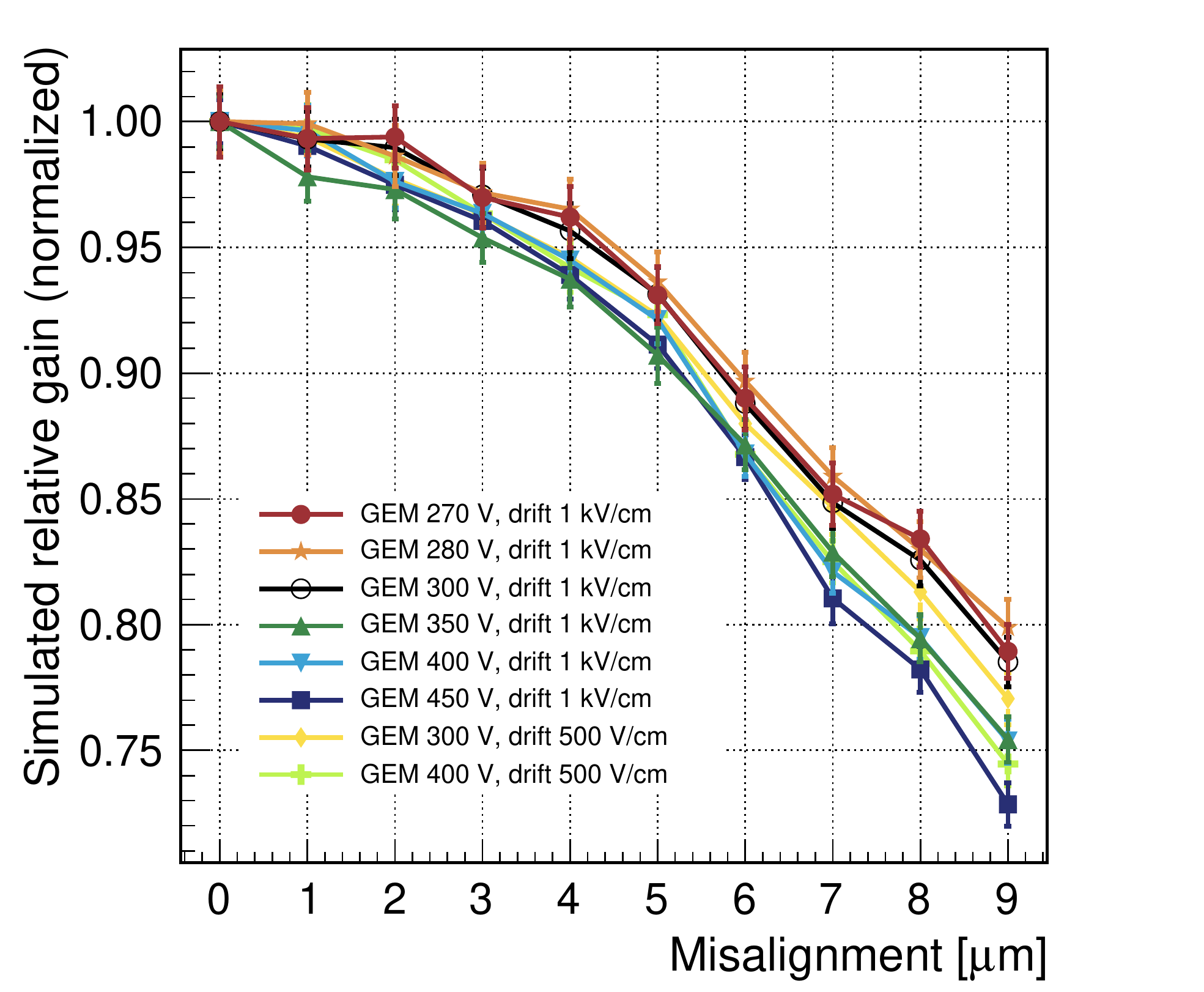}
\caption{Simulated relative gain as a function of the misalignment for standard hole geometries of 70-50-70\,$\mu$m using Garfield++. The different curves correspond to different voltages across the GEM electrodes and/or different drift fields.\label{simulationgraph}}
\end{figure}

In addition, we simulated the gas gain dependency on the misalignment of the GEM holes for different voltages on the GEM foil and two different collection (drift) field values as shown in Fig.\,\ref{simulationgraph}. For the same drift field values, the simulated relative gain drop due to misalignment increases towards higher voltages of the GEM foils.

As already mentioned, the GEM TG3 was also measured under higher voltages of 400\,V with the result of a higher non-uniformity of the relative gas gain. The simulated results support this findings.  However, to discuss its significance more refined measurements -- also at higher voltages -- and simulations would be needed.

\section{Conclusions and outlook}
We have successfully measured the misalignment of the GEM holes with a high definition optical scanning system that can help in predicting their operational relative gas gain performance. We were able to correlate the misalignment to local gas gain variations. In addition, we extracted a relative gas gain drop of $\sim$2\,\% per $\mu$m for a misalignment ranging from 1 to 8\,$\mu$m.

We emphasize again the potential of such a high definition optical scanning system for non-destructive quality control of GEM foils that allows to choose excellent GEM foils for detector assembly.  

This result may get us closer to the goal of predicting gas gain performances of full stacked GEM foil detectors prior to assembly. It is evident that gas gain variations of individual foils in a multi-layer detector are compounded, and variation extrema potentially accumulated. This accumulation of variation in the detector could be alleviated if the gain variation of the foils was known during assembly. However, we know that it is not a simple multiplication of the resulting gas gain maps as also other factors might play a crucial role, such as the charge up of the dielectrics during operation for example.

\section*{Acknowledgements}

We thank the personnel of the Detector Laboratory at the Helsinki Institute of Physics and the Wigner Research Centre for Physics that are involved in the infrastructure build-up. In addition we thank Eraldo Oliveri for his valuable help with the DAQ. This work was supported by the infrastructure grant FIRI 272921 of the Academy of Finland and by the Hungarian National Research, Development and Innovation Office (NKFIH) under the contract numbers OTKA K135515 and the NKFIH grant 2019-2.1.6-NEMZKI-2019-00011.

\clearpage
\newpage

\appendix
\section{Feature maps of the GEM foils under test}\label{tgfeatures}

\subsection{Foil TG1}
\begin{centering}
\includegraphics[height=5.8cm]{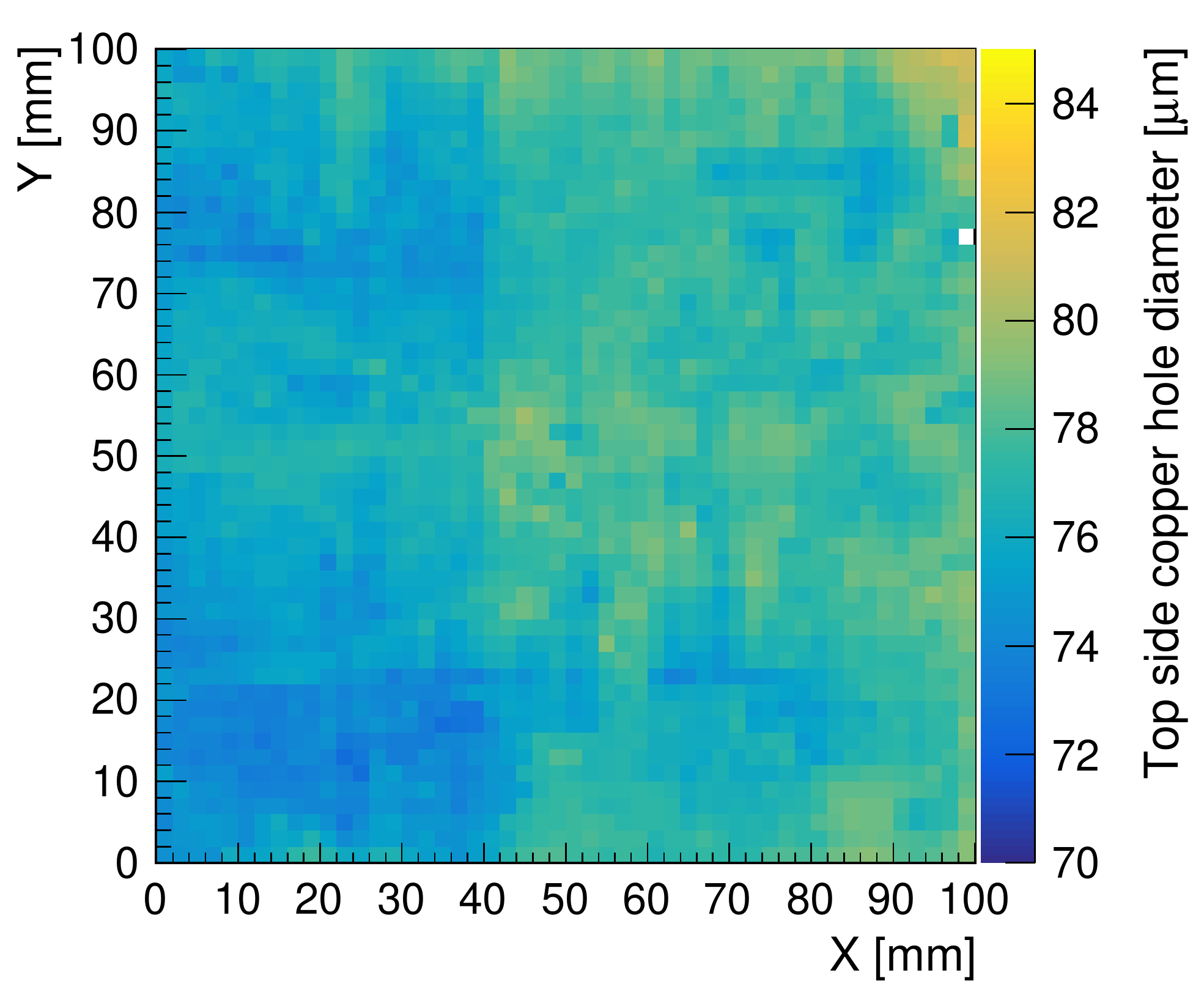}
\captionof{figure}{Top side copper hole diameter map of foil TG1.}
\end{centering}
\vspace{1em}
\begin{centering}
\includegraphics[height=5.8cm]{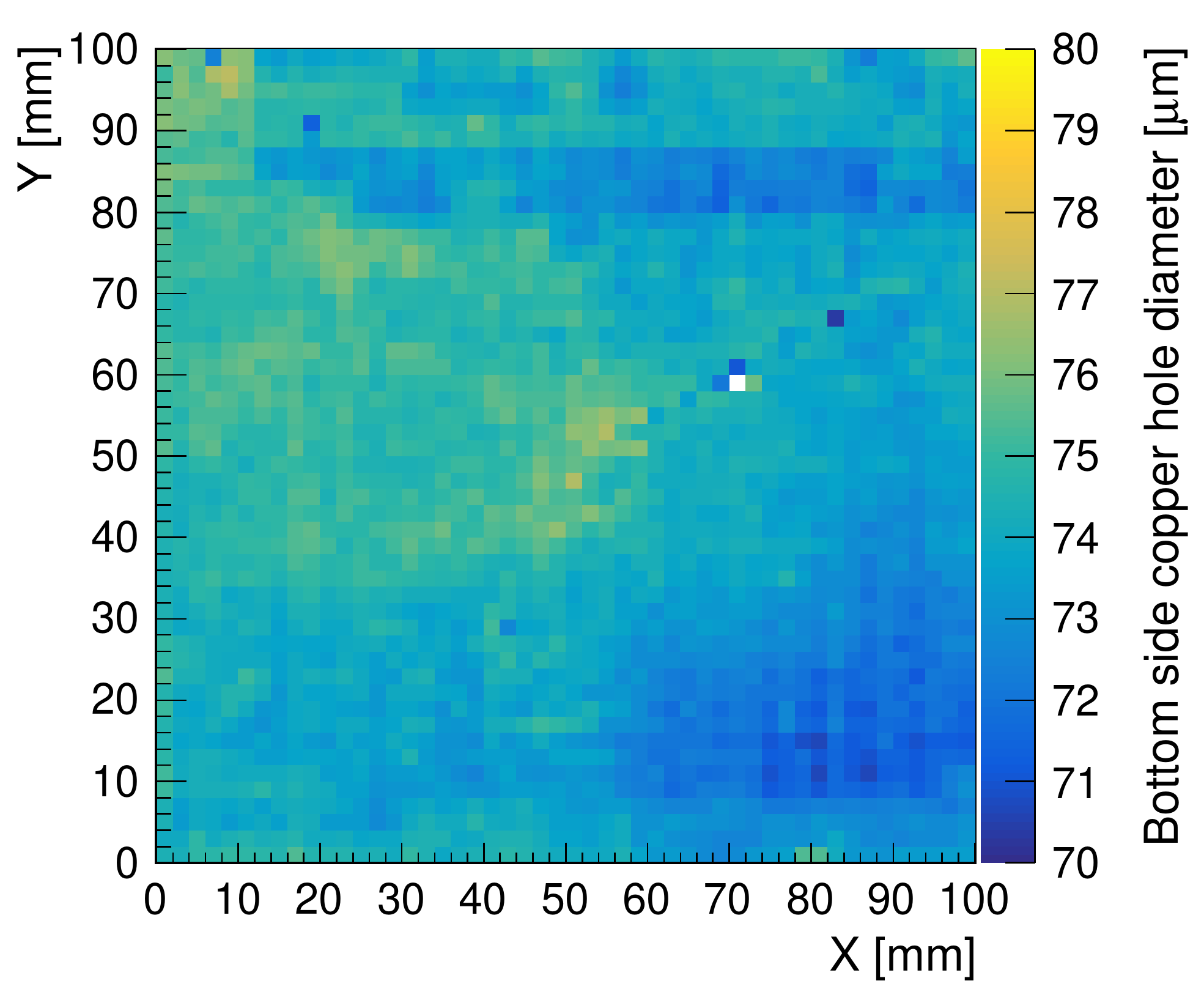}
\captionof{figure}{Bottom side copper hole diameter map of foil TG1.}
\end{centering}
\vspace{1em}
\begin{centering}
\includegraphics[height=5.8cm]{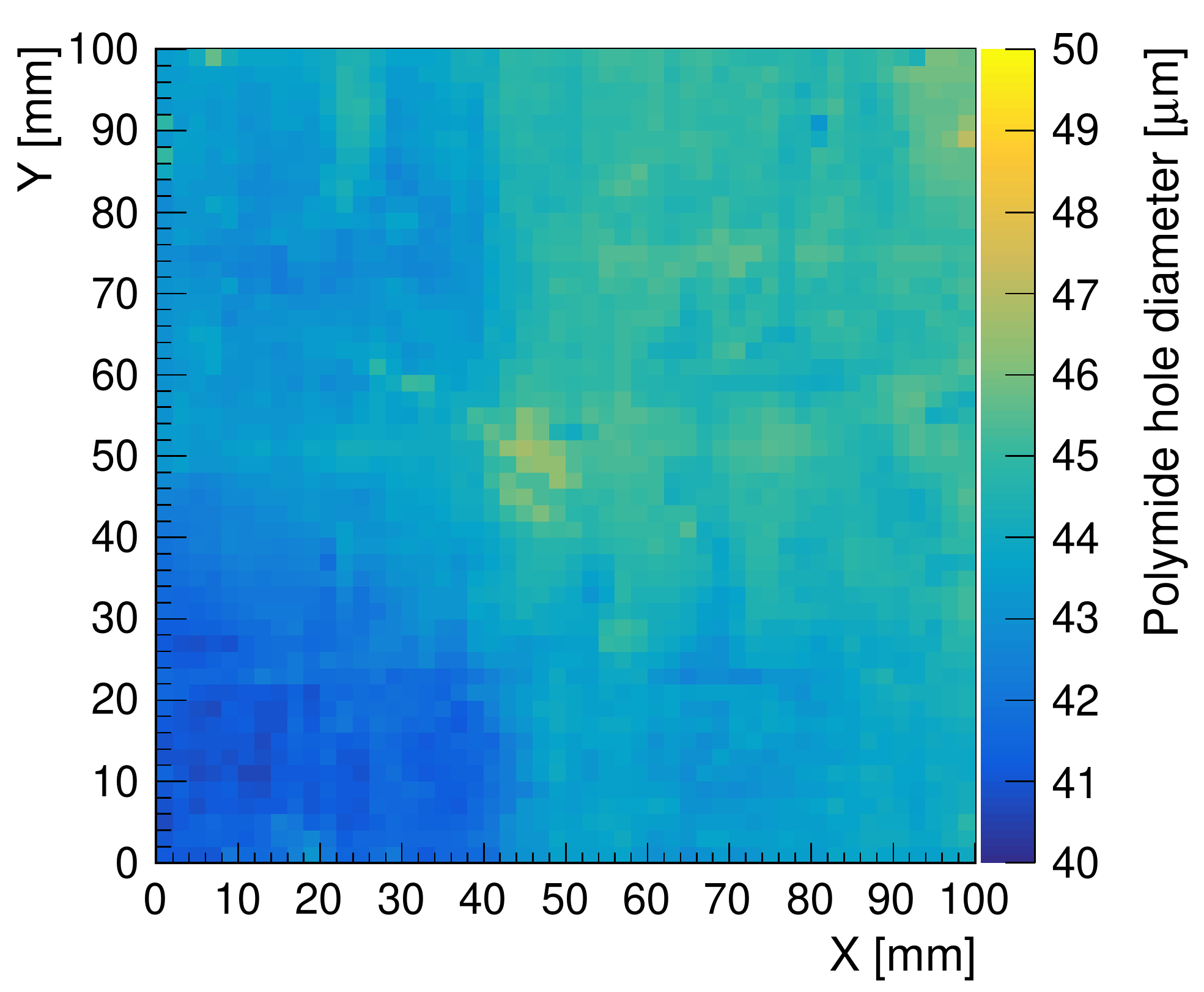}
\captionof{figure}{Polyimide hole diameter map of foil TG1.}
\end{centering}
\vspace{1em}
\begin{centering}
\includegraphics[height=5.8cm]{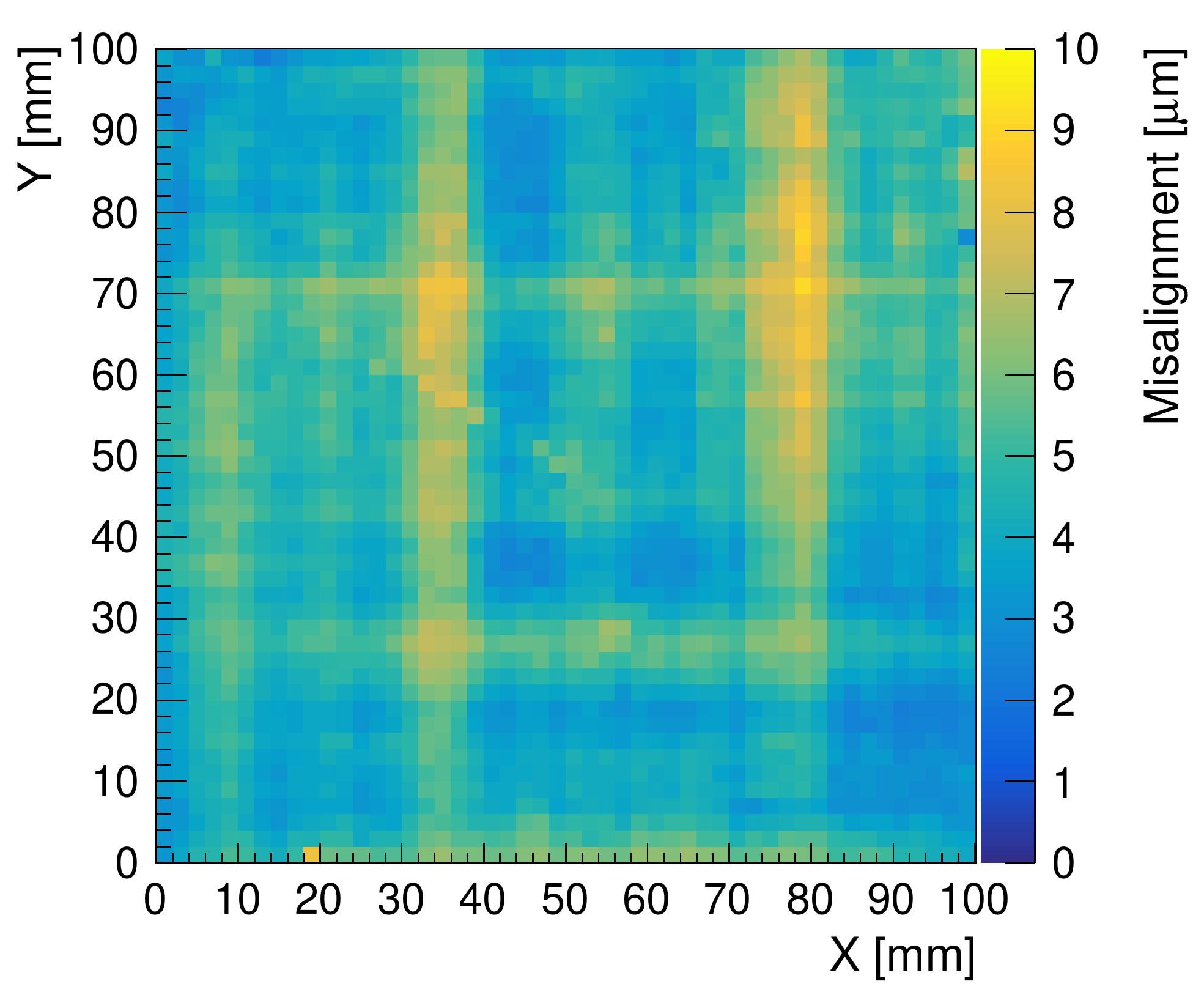}
\captionof{figure}{Misalignment map of foil TG1.}
\end{centering}
\vspace{1em}
\begin{centering}
\includegraphics[height=5.8cm]{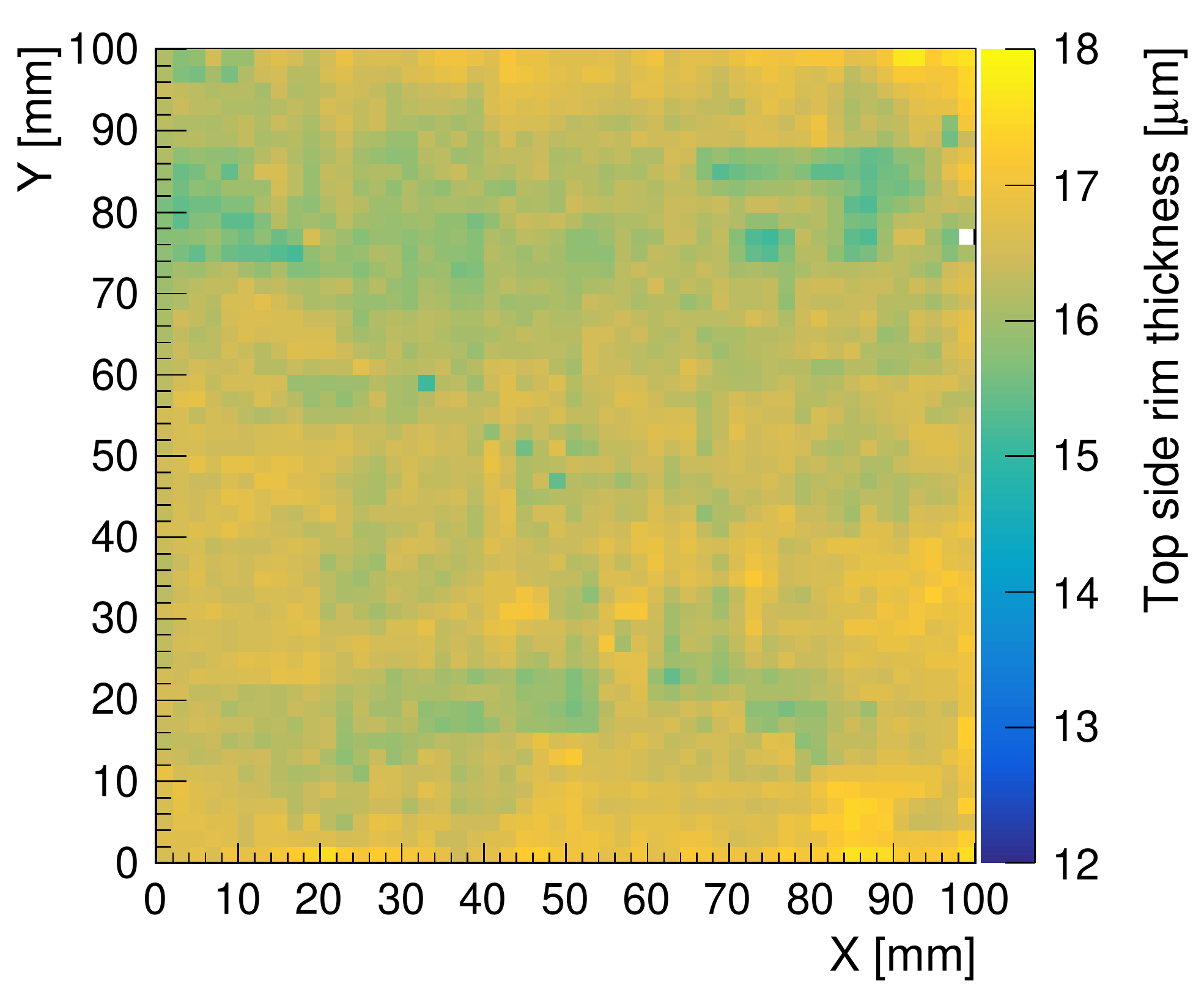}
\captionof{figure}{Top side polyimide rim thickness projected to the x-y plane of foil TG1.}
\end{centering}
\vspace{1em}
\begin{centering}
\includegraphics[height=5.8cm]{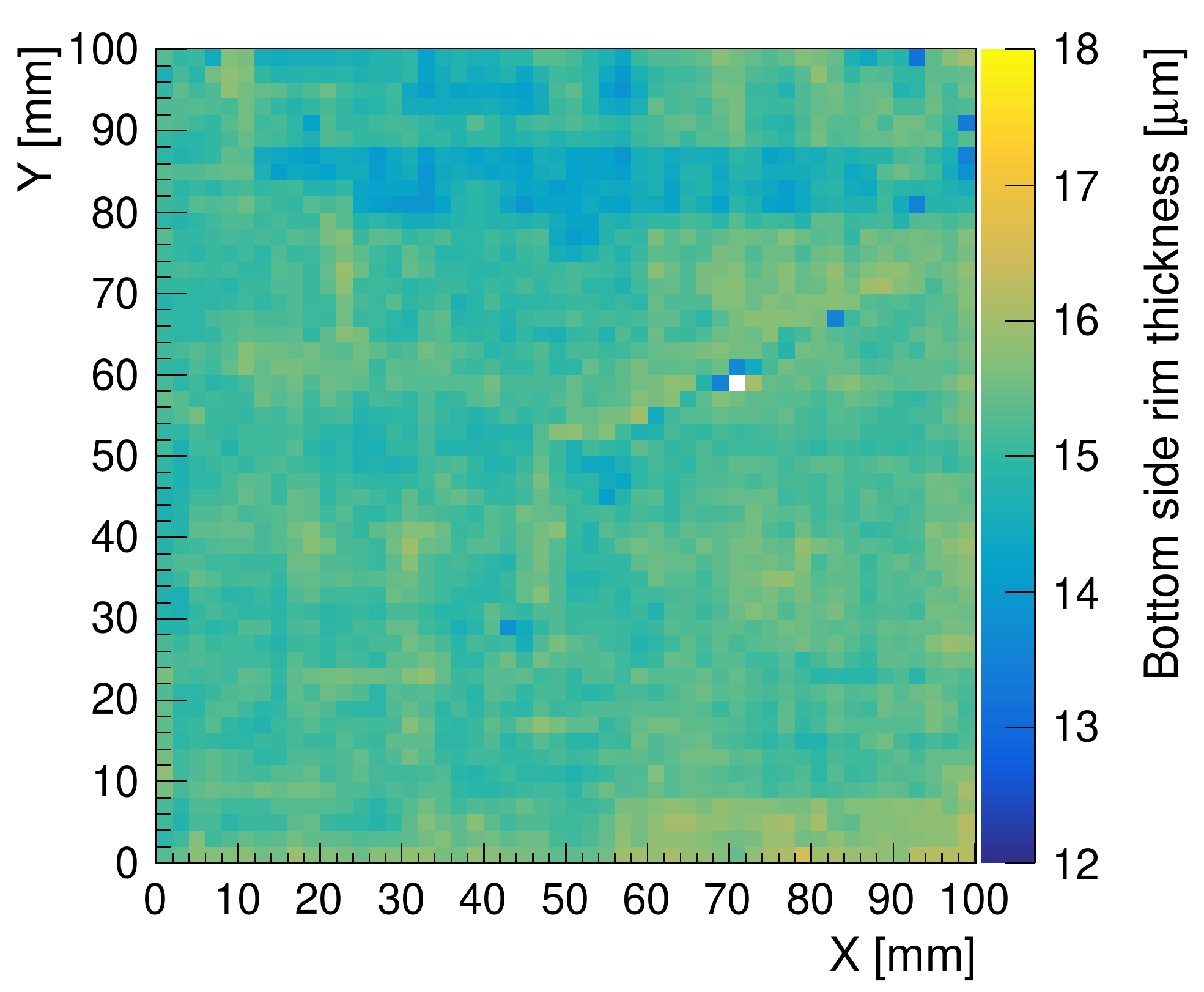}
\captionof{figure}{Bottom side polyimide rim thickness projected to the x-y plane of foil TG1.}
\end{centering}
\newpage
\subsection{Foil TG2}
\begin{centering}
\includegraphics[height=6.cm]{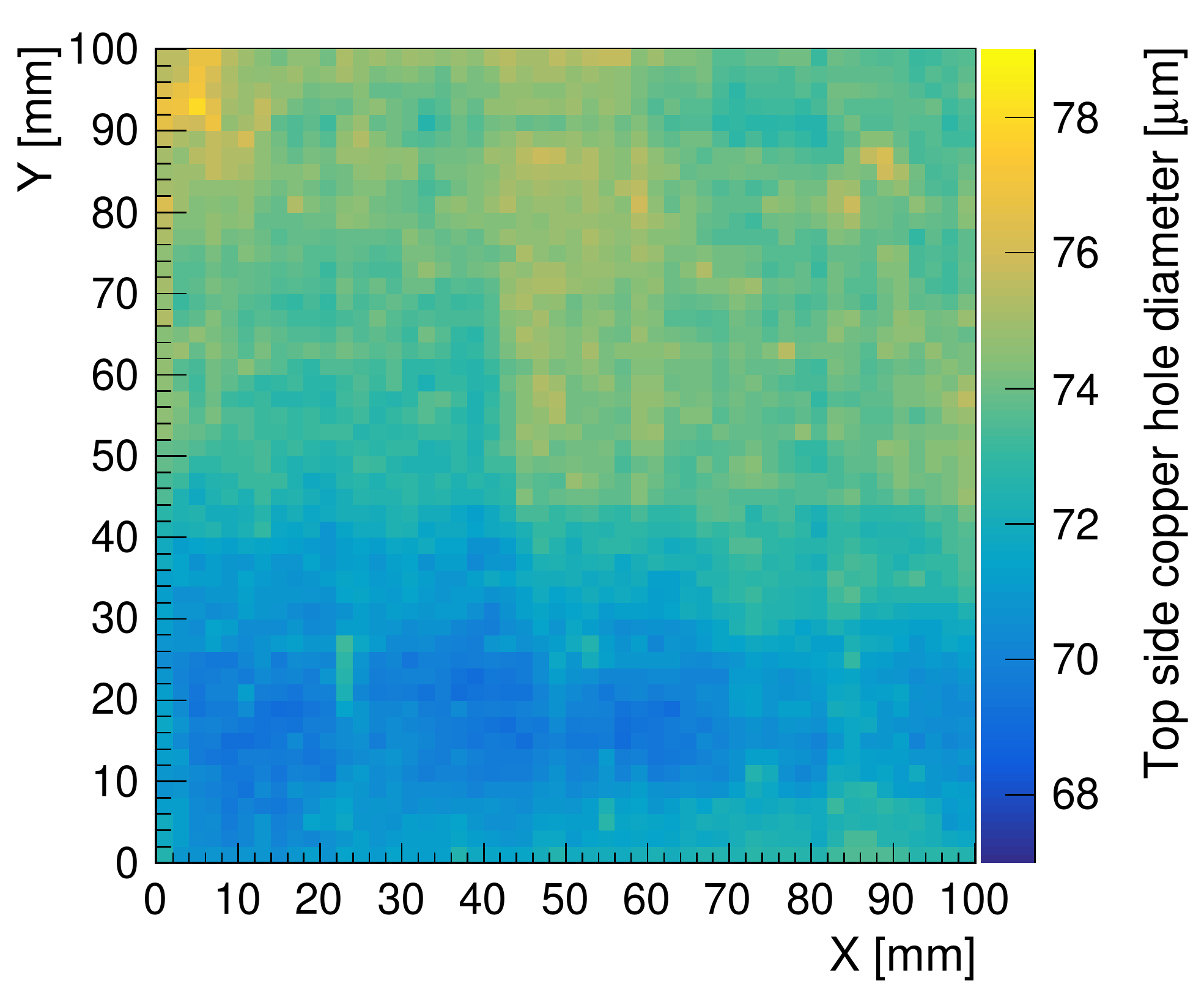}
\captionof{figure}{Top side copper hole diameter map of foil TG2.}
\end{centering}
\vspace{1em}
\begin{centering}
\includegraphics[height=6.cm]{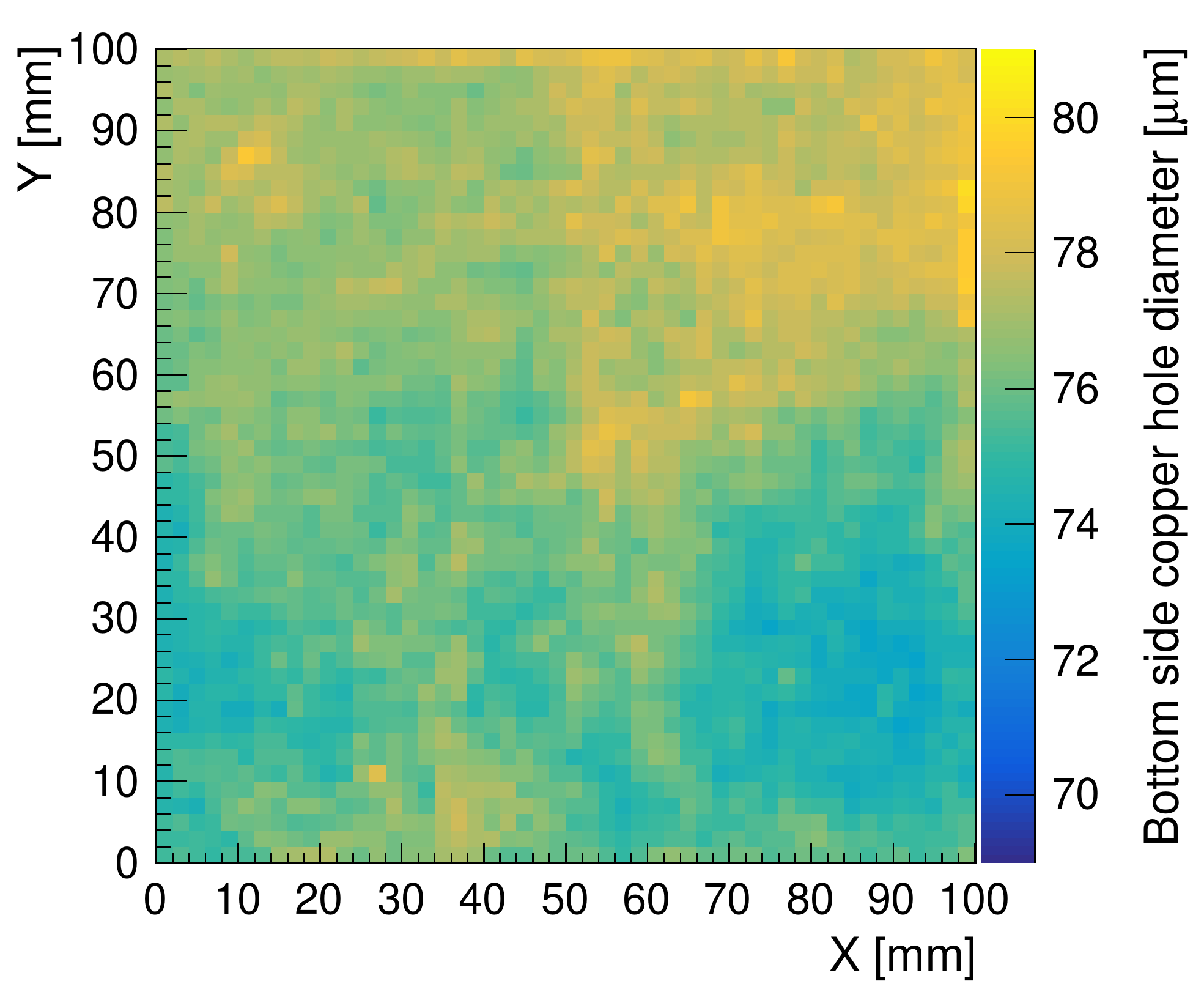}
\captionof{figure}{Bottom side copper hole diameter map of foil TG2.}
\end{centering}
\vspace{1em}
\begin{centering}
\includegraphics[height=6.cm]{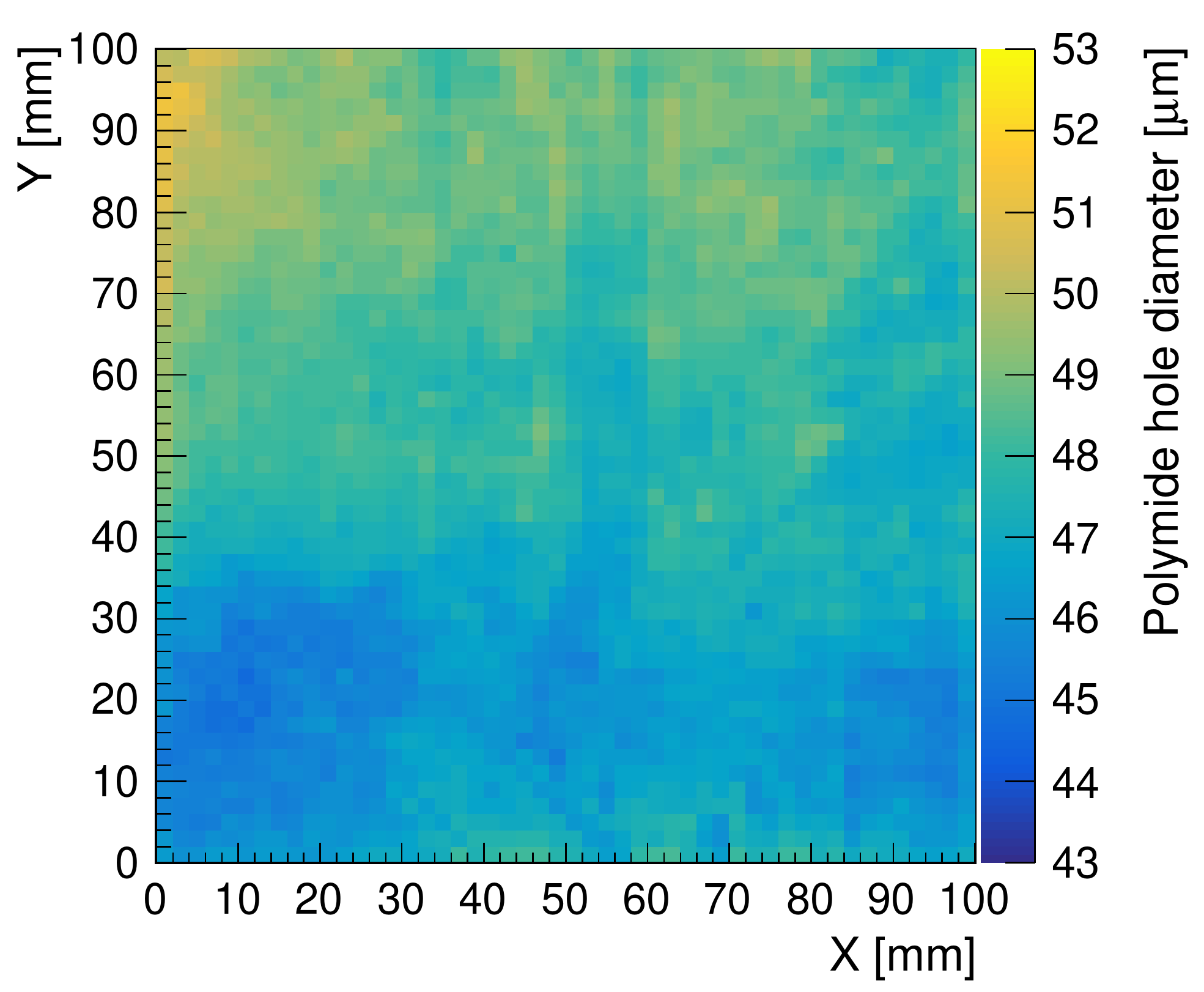}
\captionof{figure}{Polyimide hole diameter map of foil TG2.}
\end{centering}
\vspace{1em}
\begin{centering}
\includegraphics[height=6.cm]{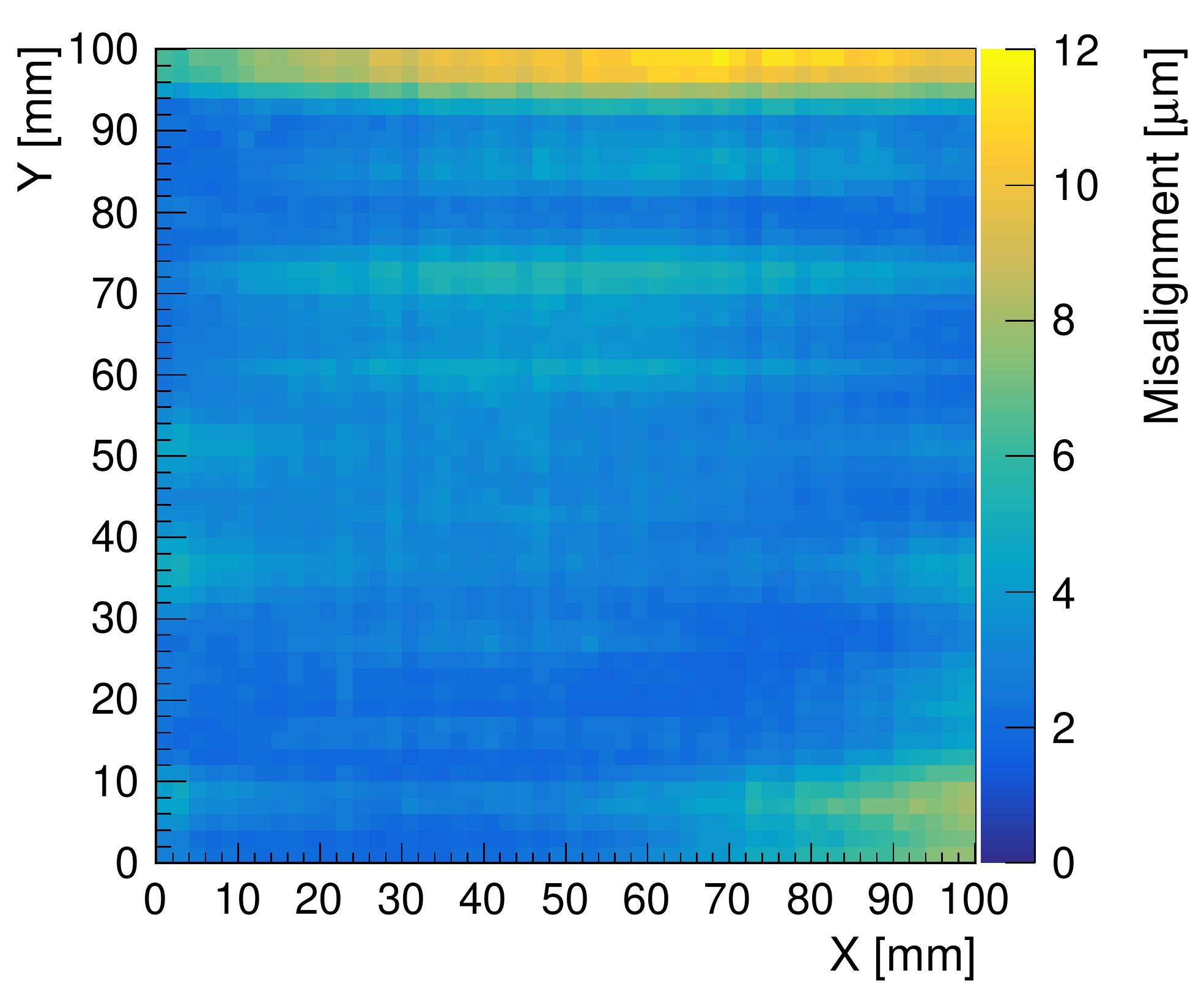}
\captionof{figure}{Misalignment map of foil TG2.}
\end{centering}
\vspace{1em}
\begin{centering}
\includegraphics[height=6.cm]{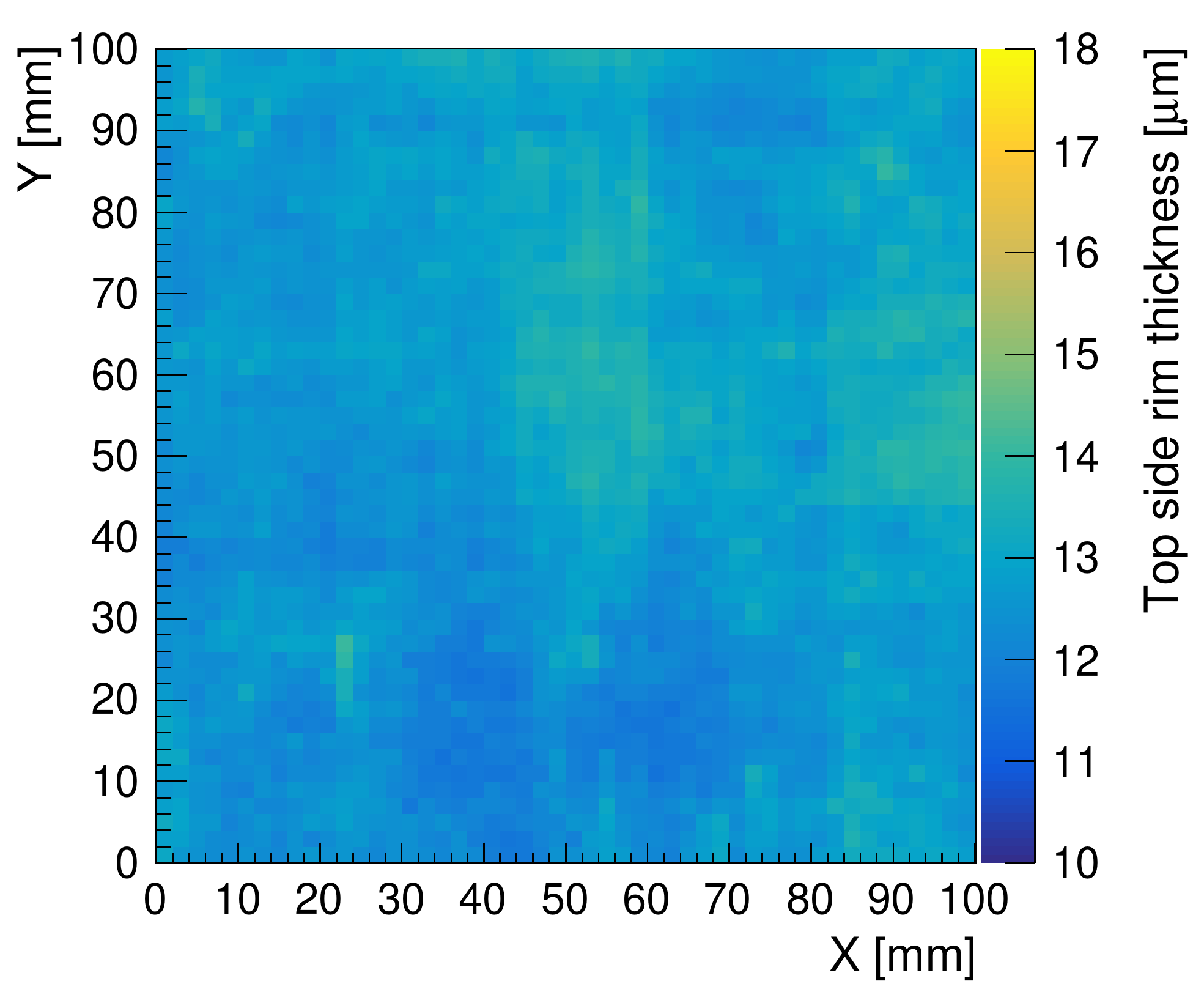}
\captionof{figure}{Top side polyimide rim thickness projected to the x-y plane of foil TG2.}
\end{centering}
\vspace{1em}
\begin{centering}
\includegraphics[height=6.cm]{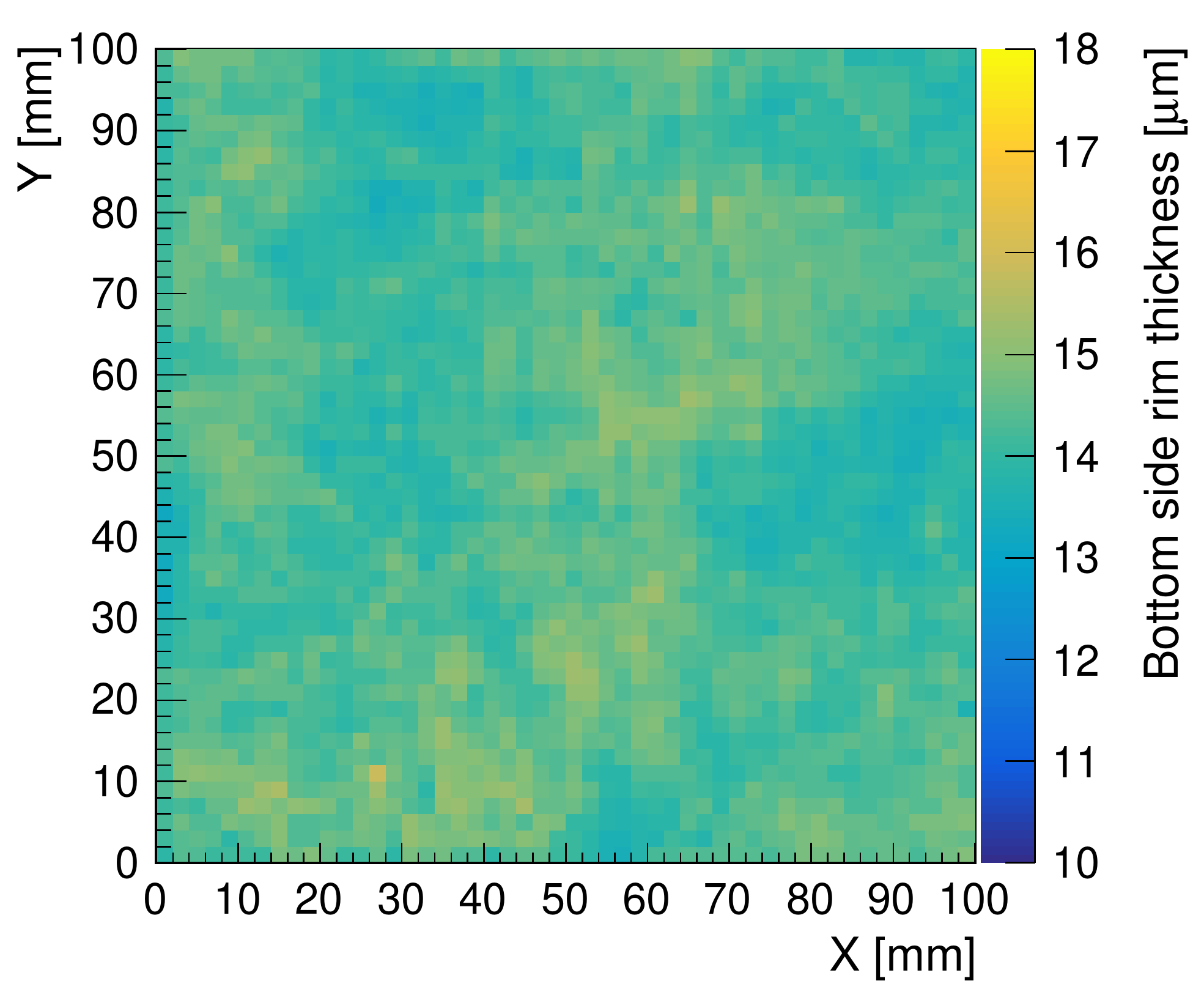}
\captionof{figure}{Bottom side polyimide rim thickness projected to the x-y plane of foil TG2.}
\end{centering}
\newpage
\subsection{Foil TG3}
\begin{centering}
\includegraphics[height=6.cm]{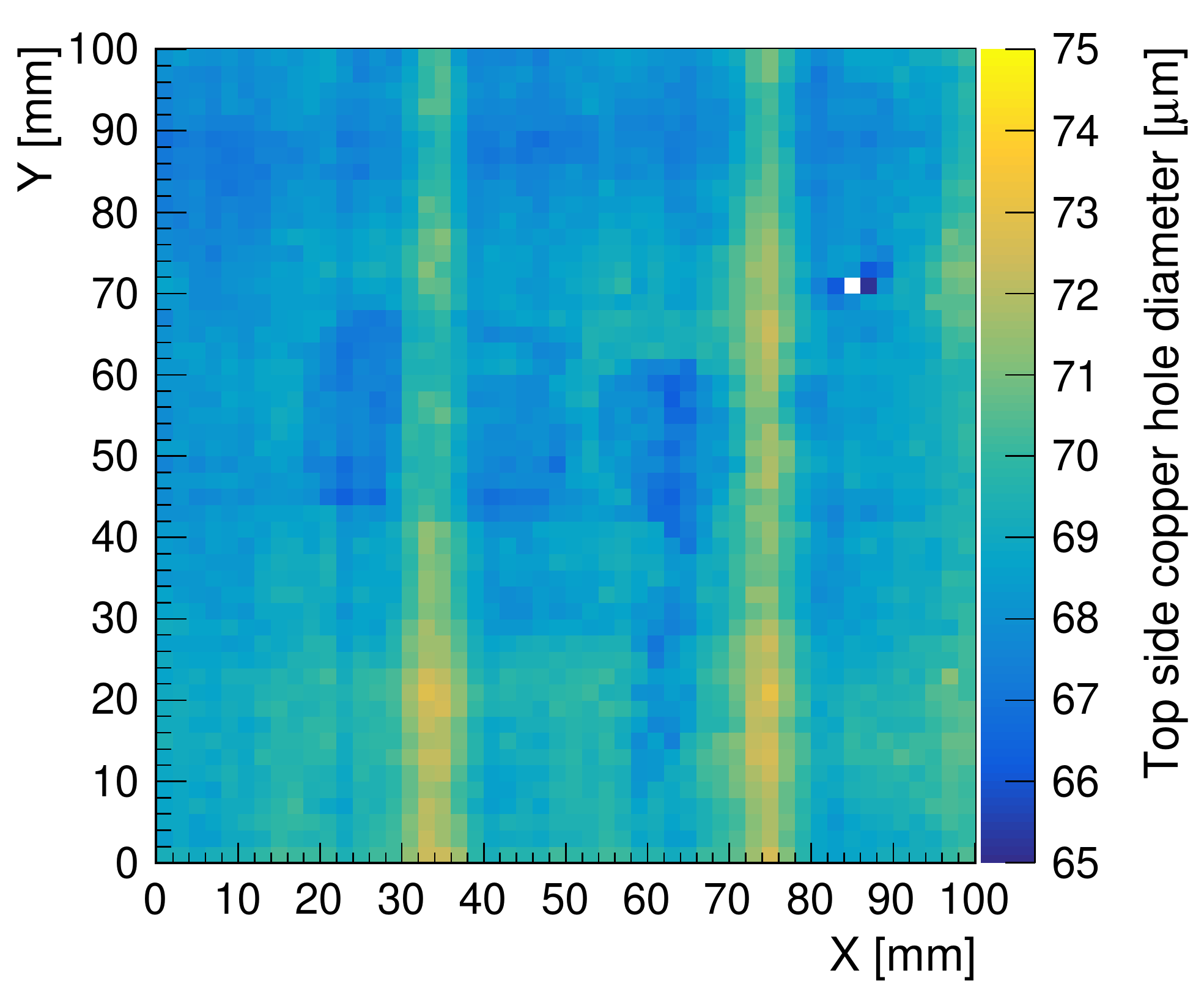}
\captionof{figure}{Top side copper hole diameter map of foil TG3.}
\end{centering}
\vspace{1em}
\begin{centering}
\includegraphics[height=6.cm]{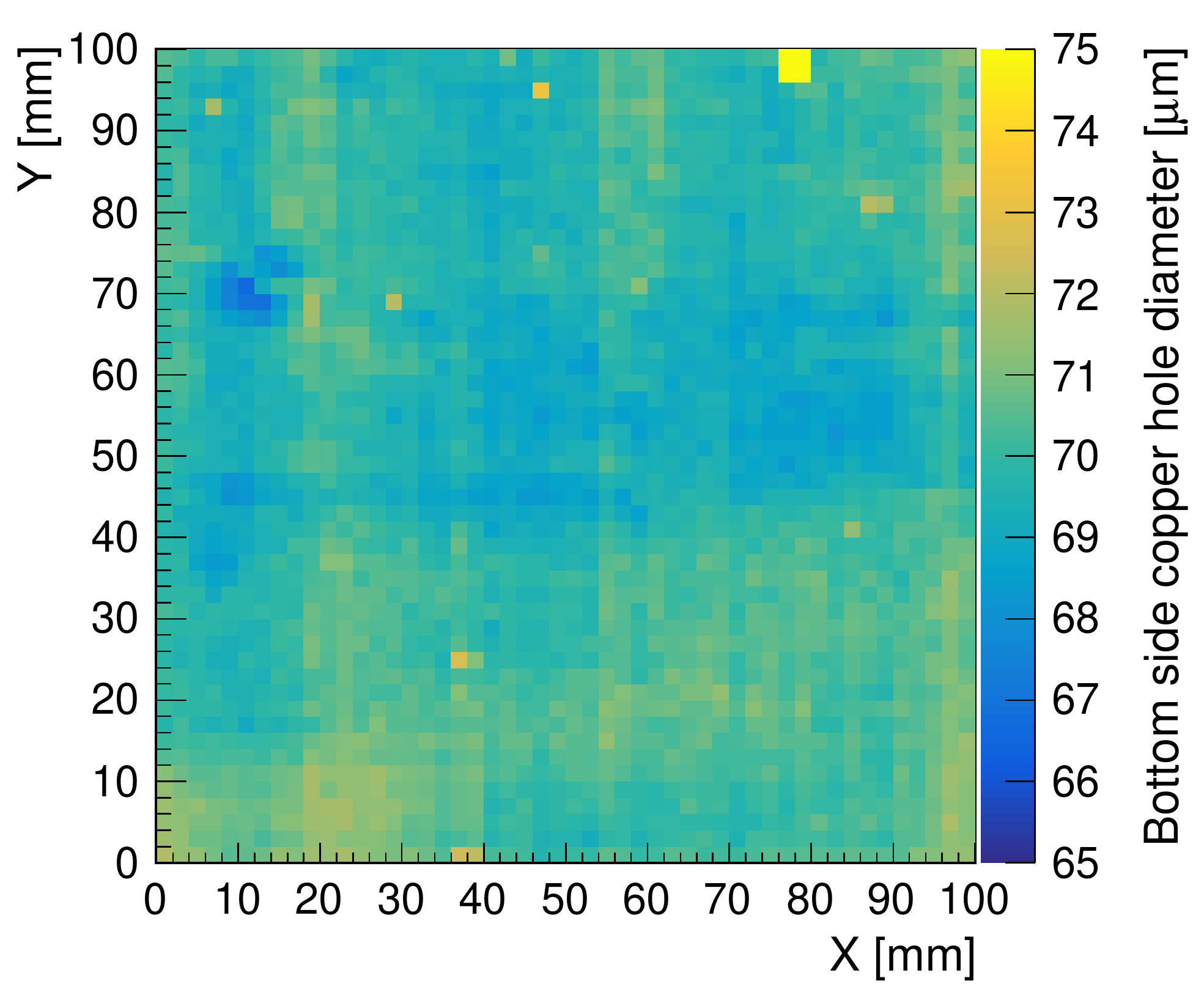}
\captionof{figure}{Bottom side copper hole diameter map of foil TG3.}
\end{centering}
\vspace{1em}
\begin{centering}
\includegraphics[height=6.cm]{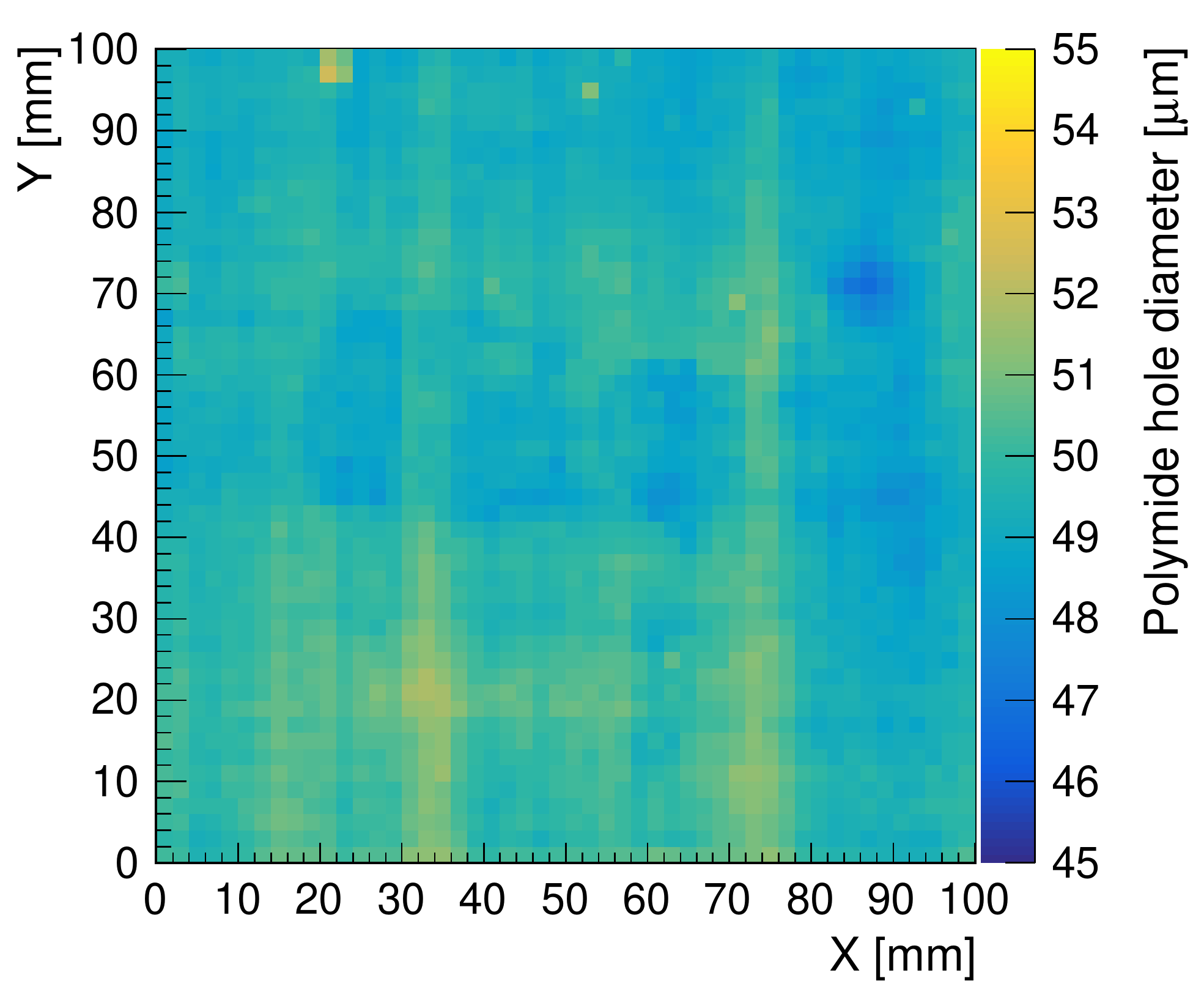}
\captionof{figure}{Polyimide hole diameter map of foil TG3.}
\end{centering}
\vspace{1em}
\begin{centering}
\includegraphics[height=6.cm]{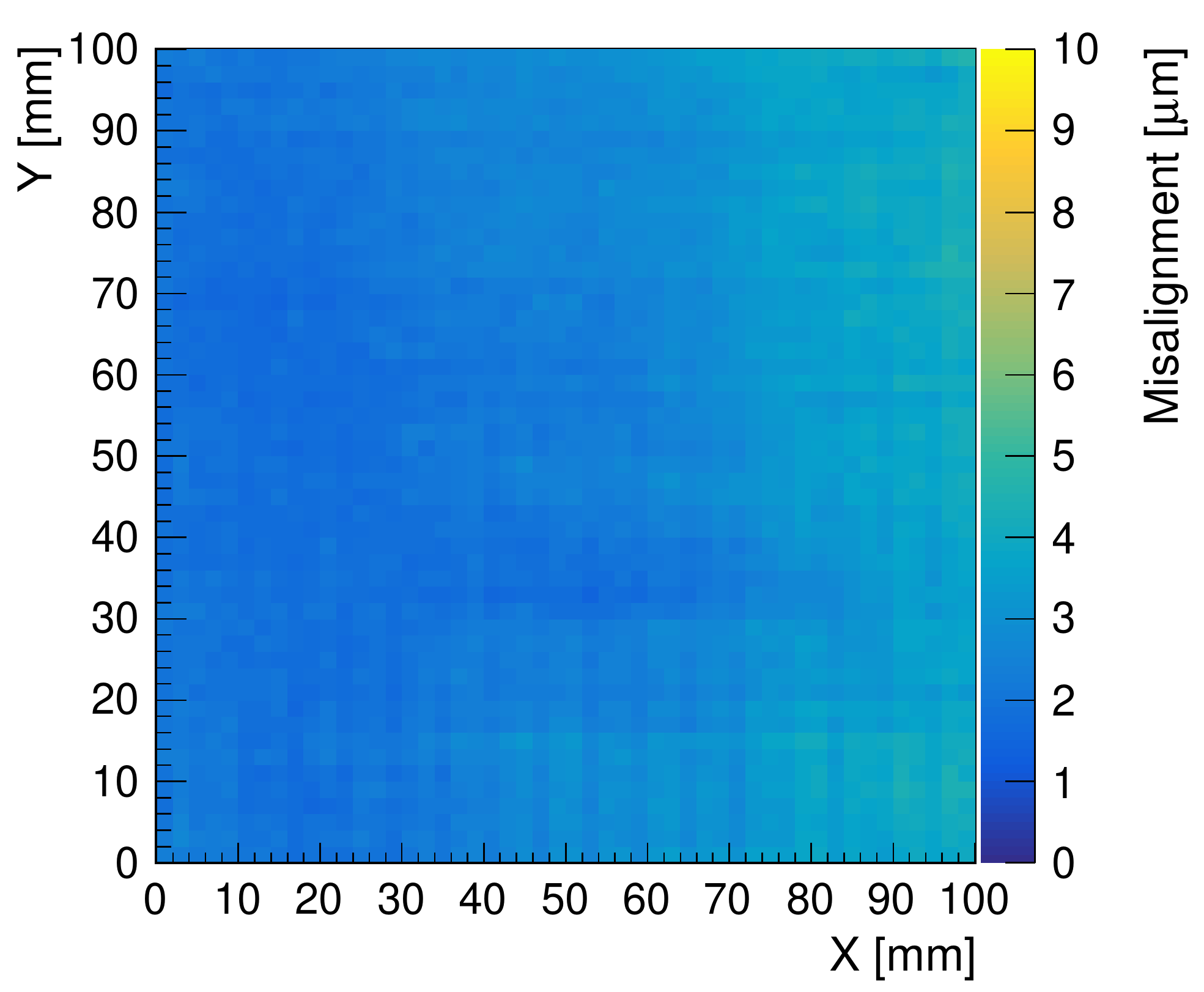}
\captionof{figure}{Misalignment map of foil TG3.\label{smooth}}
\end{centering}
\vspace{1em}
\begin{centering}
\includegraphics[height=6.cm]{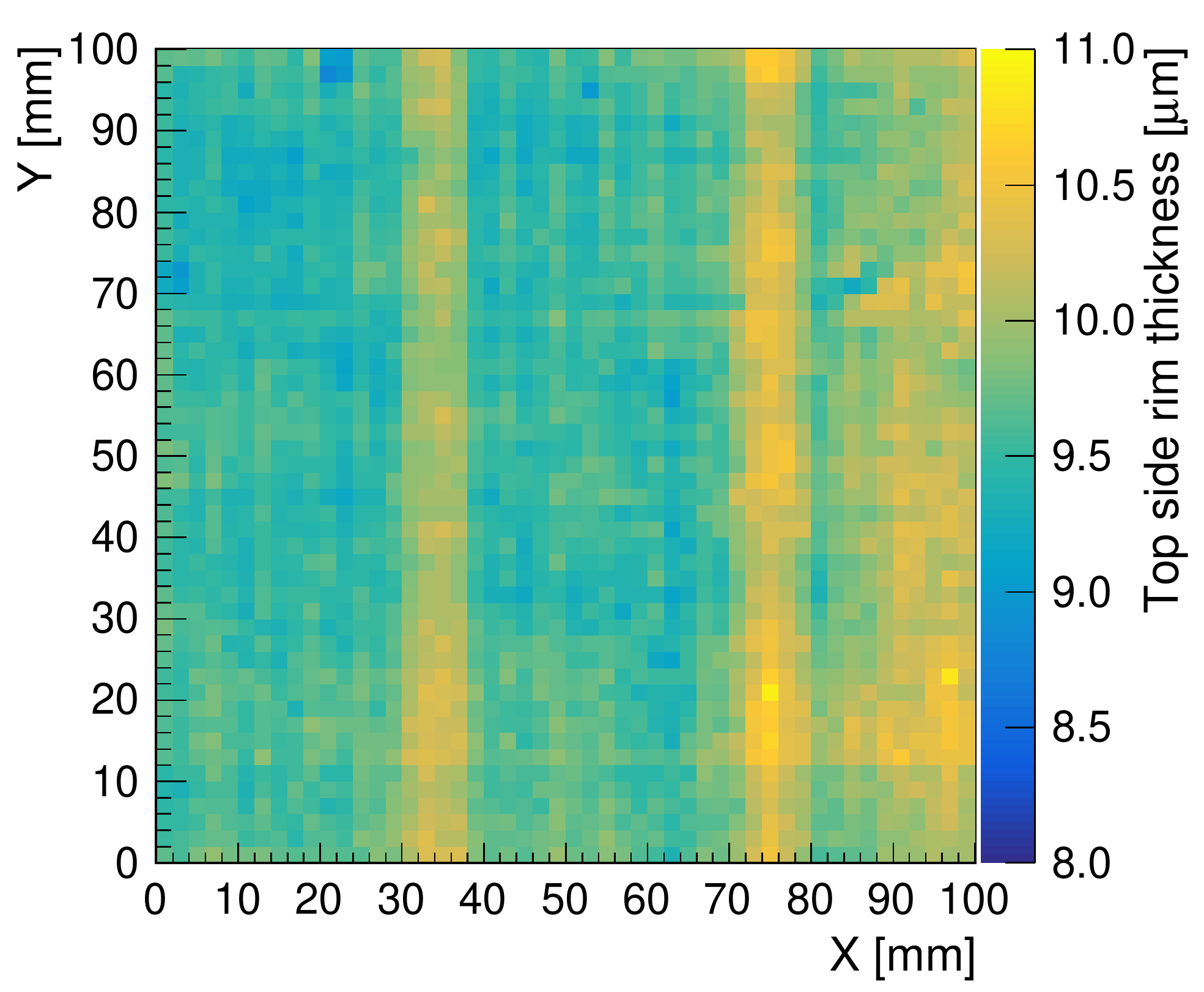}
\captionof{figure}{Top side polyimide rim thickness projected to the x-y plane of foil TG3.}
\end{centering}
\vspace{1em}
\begin{centering}
\includegraphics[height=6.cm]{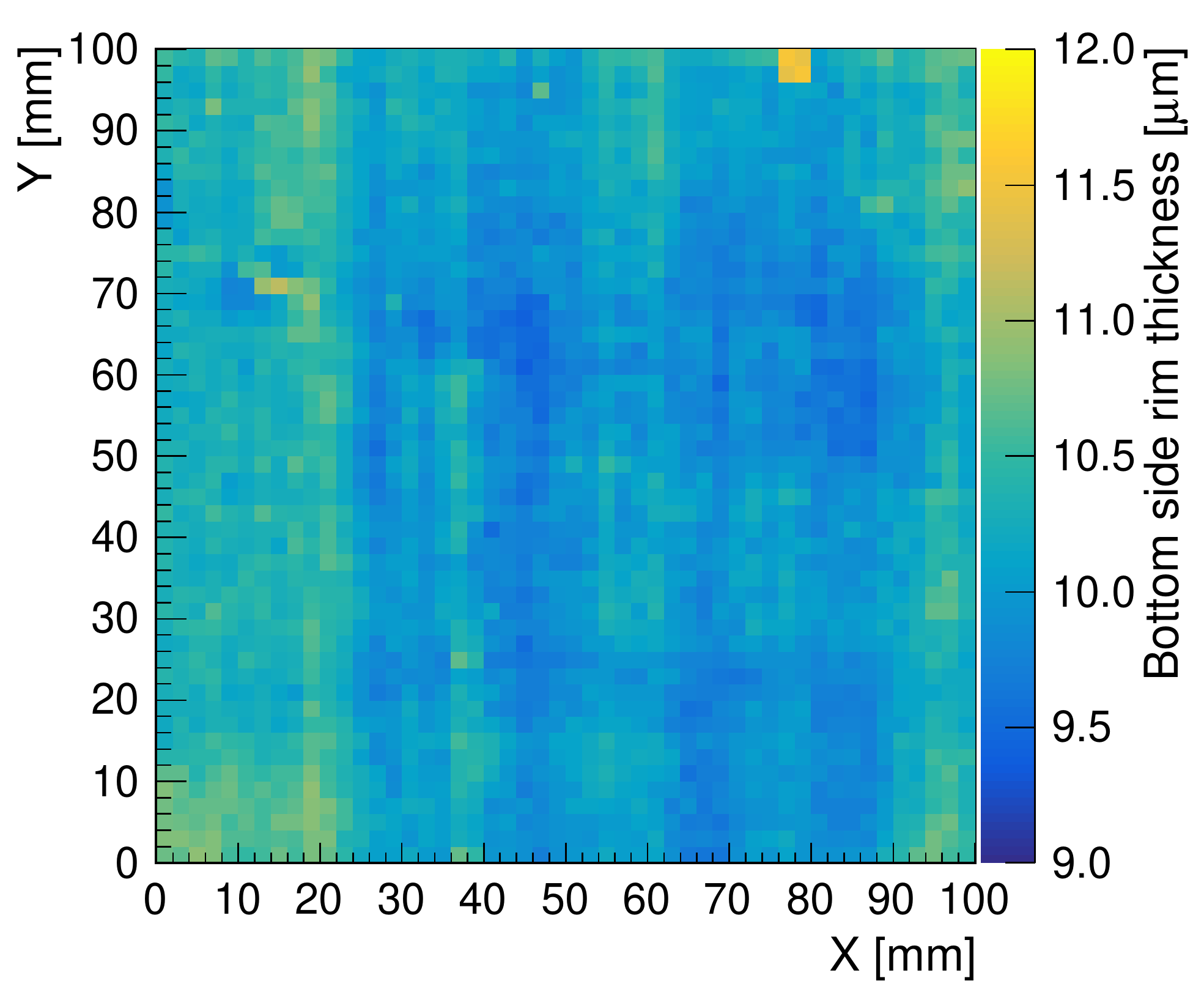}
\captionof{figure}{Bottom side polyimide rim thickness projected to the x-y plane of foil TG3.}
\end{centering}

\newpage

\section{GEM foil orientation}\label{GEMorientation}

Here we define the orientation of the GEM foils as used in our measurements. We refer to the \textit{top side} in case the foil is installed in the measurement systems (same case for all setups) in such a way that the supply contacts for the electrodes are on the lower right side as shown in Fig.\,\ref{GEMorientation}. We refer to the \textit{bottom side} when the foils in its orientation as in Fig.\,\ref{GEMorientation} is flipped 180 degree along the vertical axis such that the supply electrodes are on the lower left side.
\vspace{1em}

\begin{centering}
\includegraphics[height=6.cm]{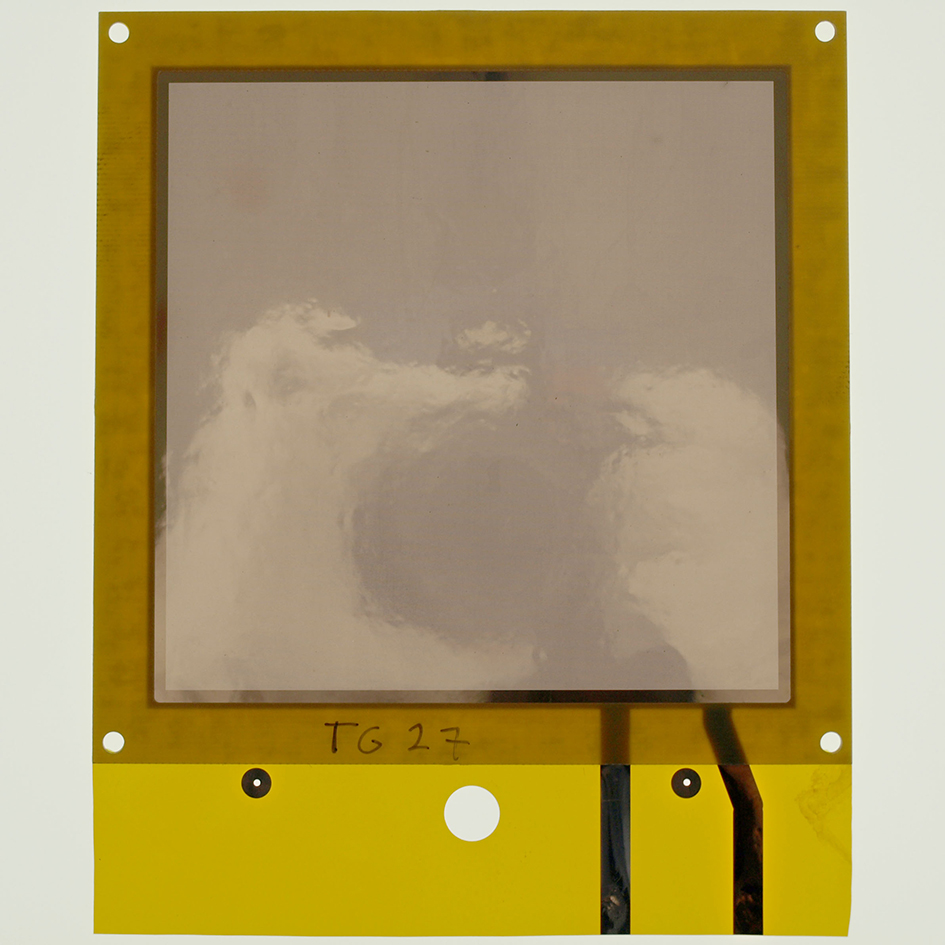}
\captionof{figure}{Image of a GEM foil in its self-defined top side orientation.\label{GEMorientation}}
\end{centering}

\end{document}